\RequirePackage{fix-cm}
\documentclass[twocolumn,epjc3]{svjour3}  
\smartqed  
\RequirePackage{graphicx}

\RequirePackage{ae}
\RequirePackage{bm} 
\RequirePackage{color}
\RequirePackage{amssymb}
\RequirePackage{amsmath}
\RequirePackage{amsfonts}
\RequirePackage{mathtools}
\RequirePackage{graphicx}
\RequirePackage{slashed}
\RequirePackage{wasysym}
\RequirePackage{setspace}
\RequirePackage{subfigure}
\RequirePackage{ulem}
\RequirePackage{placeins}

\RequirePackage{tabularx}
\RequirePackage{algpseudocode}
\RequirePackage{nicefrac}
\RequirePackage{bbm}
\RequirePackage{calligra}
\RequirePackage{microtype}

\usepackage{cite}
\newcommand{\dns}[1]{} 

\newcommand*{\I}{ {\rm i} }
\newcommand*{\ee}{ {\rm e} }
\newcommand*{\kk}{ {\rm k} }
\DeclareMathAlphabet{\mathcalligra}{T1}{calligra}{m}{n}

\newlength{\figlen}
\newlength{\figlenB}
\newlength{\figp}
\newcount\Bool	
\def\eq$#1${\begin{multline}#1\end{multline}}
\newcommand{\br}{\\}

\newcommand{\txt}[2]{
  \ifnum\Bool=1 {#1} 
  \else	{#2}
  \fi
}

\newcommand{\ck}[1]{\textcolor{black}{#1}}

\journalname{The European Physical Journal Plus}
\begin{document} \sloppy

\title{Phase-space analysis of the Schwinger effect in inhomogeneous electromagnetic fields}

\author{Christian Kohlf\"urst \thanksref{e1,addr1,addr2}}

\thankstext{e1}{e-mail: C.Kohlfuerst@gsi.de}


\institute{\emph{Helmholtz-Institut Jena, Fr\"obelstieg 3, 07743 Jena, Germany} \label{addr1}
           \and
           Theoretisch-Physikalisches Institut, Abbe Center of Photonics, \\
Friedrich-Schiller-Universit\"at Jena, Max-Wien-Platz 1, 07743 Jena, Germany \label{addr2}
}

\date{}

\maketitle

\begin{abstract}
Schwinger pair production in spatially and temporally inhomogeneous electric and magnetic fields is studied. 
The focus is on the particle phase-space distribution within a high-intensity few-cycle pulse. 
Accurate numerical solutions of a quantum kinetic theory (DHW formalism) are presented in momentum space and, with the aid of
coarse-graining techniques, in a mixed spatial-momentum representation.
Additionally, signatures of the carrier-envelope phase as well as spin-field interactions are discussed on the basis of a trajectory-based model
taking into account instantaneous pair production and relativistic single-particle dynamics. Although our simple semi-classical single-particle model
cannot describe every aspect of the particle production process (quantum interferences), essential features such as spin-field interactions are captured. 

\end{abstract}

\section{Introduction}

The creation of matter via light is one of the most striking features of strong-field QED \cite{Schwinger:1951nm,Heisenberg:1935qt,Sauter:1931zz}.
Even though multiphoton pair production has been measured experimentally \cite{Burke:1997ewA,Burke:1997ewB},
the prominent Schwinger effect still waits for an experimental verification \cite{Heinzl:2008an,Ringwald:2001ib}. The recent effort, however,
that is put into the research field, c.f. upcoming laser facilities \cite{LasersA,LasersB}, could bring it closer to detection \cite{Marklund:2008gj,mStar:FELsA,mStar:FELsB,Turcu:2015cca}.

Due to the advent of lasers that can probe the relativistic regime, there has been a substantial activity in studying strong-field QED in recent 
years \cite{Dunne:2004nc,reviewA,reviewB,reviewC,reviewD,reviewE}.
With the recent advances of a measurement of light-by-light scattering of quasi-real photons \cite{LightA,LightB} the research field is expected to attract even more attention.
In this regard, Schwinger pair production represents the perfect show case, because it is a non-perturbative effect that inevitably unites the highly relativistic regime with the quantum regime.  

Moreover, theoretical approaches that have been initially developed
in the last century, improved substantially in the last twenty years. In turn, this progress
paved the way for investigations that simply were not possible with previous techniques \cite{Hebenstreit:2011wk}. Especially the
broad usage of contemporary numerical methods is worth mentioning here, because it enabled to establish kinetic theories \cite{Kluger:1992md,Rafelski:1993uh,Vasak:1987umA,Vasak:1987umB,BB,Smolyansky:1997fcA,Smolyansky:1997fcB,Smolyansky:1997fcC,Smolyansky:1997fcD}.

In order to lay a solid foundation for the discussion on the Schwinger effect we employ the Dirac-Heisenberg-Wigner (DHW) formalism \cite{Hebenstreit:2008ae,Vasak:1987umA,Vasak:1987umB,BB}. 
The main advantage of the DHW formalism is that it automatically combines quantum electrodynamics with notions familiar from statistical physics \cite{Akkermans:2011yn,Hebenstreit:2009km}. 
Its versatility allows to incorporate temporal \cite{Smolyansky:1997fcA,Smolyansky:1997fcB,Smolyansky:1997fcC,Smolyansky:1997fcD,Hebenstreit:2009km,TempB,TempC,Blinne} as well as spatial inhomogeneities \cite{Kohlfurst:2015zxi,Hebenstreit:2011wk,Berenyi:2013eia,Kohlfurst:2015niu}. 
On top of that, the DHW formalism gives access to the complete phase-space distribution of the created particles. 
In the present work, we take advantage of this feature to
calculate various particle distributions without being limited to a momentum or a spatial representation.
Nevertheless, we also search for characteristic signatures in the particle momentum spectrum.

Performing simulations for the complete phase-space includes solving a coupled system of partial differential equations incorporating, at least in principle, an infinite series of differential operators.
Hence, to perform this ambitious task advanced numerical methods are required. In Sec. \ref{ch:Theo} we shortly summarize the most important aspects of the DHW formalism and state the equations of motion. 
We proceed by presenting the solution strategies applied to the problem, where the focus is on providing a detailed insight into the technical implementation, see Sec. \ref{ch:Sol}.
In Sec. \ref{ch:CG} we introduce coarse-graining methods and discuss their advantage when analyzing phase-space quasiprobabilities.

In addition, we perform a comparison with an effective theory for the particle production rate in Sec. \ref{ch:Traj}. 
The creation of particles is investigated in analogy to the formation of quark pairs via constant chromoelectric fields, cf. the flux-tube model in Refs. \cite{Casher:1979gwA,Casher:1979gwB}.
As a result of further particle interactions with the external field, we obtain an approximate distribution function in phase-space. The big advantage of this approach is that
we get an analytic estimate for the production rate and a direct access to the particle dynamics as we can switch on/off any forces and interaction terms. To put it simple, it gives the
opportunity to understand the outcome of a complex quantum field theoretical investigation on the basis of a semi-classical model.

Furthermore, this comparison facilitates to acquire a more comprehensive picture of the Schwinger effect accompanied by the introduction of a comparatively simple model for the field. The pair production
process is then studied and interpreted on the basis of a phase-space approach and a semi-classical effective action approach.
Despite the simplicity of the employed electric and magnetic fields, nontrivial particle distributions emerge, which, in turn, can be well understood in terms of a semi-classical picture. 

We exemplarily calculate the phase-space distribution of electrons and positrons in a mixed spatial-momentum representation in Sec. \ref{ch:Coarse}. We demonstrate, that
by applying coarse-graining techniques the interpretability of the data can be greatly improved. In Sec. \ref{ch:MomSpace} we thoroughly examine particle distributions in momentum space. 
The focus is on comparing strongly inhomogeneous fields with spatially nearly homogeneous fields. We further study the influence of strong magnetic fields on the particle distribution, see 
Sec. \ref{ch:Phase}. Simultaneously, we investigate the impact of the carrier envelope phase of few-cycle pulses analyzing various field configurations and discussing the emerging interference patterns.  

\section{Schwinger pair production}

The focus of this work is entirely on Schwinger pair production; separating virtual charged particles with the aid of an electric field to create real matter \cite{Dunne:2004nc,Sauter:1931zz}.
The relevant scale of the process is given by the mass 
of the participating particles; Compton time and Compton length of an electron are $1/m_e \approx 10^{-21}$ s and $1/m_e \approx 10^{-12}$ m, respectively. 
Moreover, as the field has to provide the rest energy of the particle-antiparticle pair, the field strengths needed are of the order of
\begin{equation}
 E_{cr} = \frac{m_e^2}{e} \approx 1.3 \times 10^{16} \, \mathrm{V/cm} .
\end{equation}

\ck{In this regard it is utterly important, that the field frequency (and therefore the photon energy) is small in comparison to the particle 
rest mass, $E_{\gamma} \ll m_e$.}
More specifically, in order to avoid undesirable artifacts due to absorption processes (multi-photon pair production \cite{Kohlfurst:2014,Ruf:2009zz}, dynamically assisted
pair production \cite{PhysRevLett.101.130404}) we have to make sure, \ck{that the applied field (i) varies only slowly in time and (ii) a Fourier transform with respect to the fields temporal profile
does not yield strong peaks at nonzero energies. Additionally,} an investigation of Schwinger pair production within a phase-space approach using 
laser pulse lengths ($\tau_{\rm Laser} > 1$as $\approx 1000$ $m_e^{-1}$) is computationally expensive \cite{Blinne}. Hence, we introduce a model for the field, that \ck{meets the
requirements and} describes a given realistic situation reasonably well capturing all essential features.

The corresponding model of choice is given by
\begin{align}
 \mathbf A (t,z) &= \frac{\varepsilon E_{cr}}{\omega} \ \exp \left( -\frac{z^2}{\lambda^2} \right) \ \exp \left( -\frac{t^2}{\tau^2} \right) \sin \left( \omega t + \phi \right) \mathbf{e}_x. \label{equ:A}
\end{align}
Here, $\varepsilon$ determines the electric field strength, $\tau$ sets the temporal scale and $\lambda$ specifies the spatial scale. \ck{To make sure, that absorption processes
do not play a role in our simulations we only employ few-cycle pulses, $\omega \tau \approx 1$. In this regard, the parameter $\omega$ should not be confused with a photon energy, due to
the lack of a dominant field frequency.
\footnote{This is a feature of few-cycle pulses and is best seen from a Fourier transform of Eq. \eqref{equ:A} with respect to time. In the Schwinger regime ($\omega \tau \approx 1$ and $\omega \ll m_e$)
the photons span a wide range of energies.}}
Rather it should be seen as a control parameter ensuring a few-cycle pulse
and determining the ratio between electric and magnetic field strength. The parameter $\phi$ gives us control over the carrier-envelope phase.

Within this work, the electric and magnetic fields are derived from Eq.~\eqref{equ:A} reading
\begin{multline}
 \mathbf E (t,z) = -\partial_t \mathbf A (t,z) = \\
   \frac{\varepsilon E_{cr}}{\omega} \ \exp \left( -\frac{z^2}{\lambda^2} \right) \ \exp \left( -\frac{t^2}{\tau^2} \right) \hspace{2.5cm} \\
   \times \frac{\Big( 2t \sin \left(\omega t + \phi \right) -\omega \tau^2 \cos \left( \omega t + \phi \right) \Big)}{\tau^2} \ \mathbf{e}_x, \label{equ:E} 
\end{multline}
\begin{multline}
 \mathbf B (t,z) = \boldsymbol \nabla \times \mathbf A (t,z) = \\
  -\frac{\varepsilon E_{cr}}{\omega} \ \exp \left( -\frac{z^2}{\lambda^2} \right) \ \exp \left( -\frac{t^2}{\tau^2} \right) 
   \ \frac{ 2z \sin \left( \omega t + \phi \right) }{\lambda^2} \ \mathbf{e}_y. \label{equ:B}
\end{multline}

\begin{figure}[t]
      \begin{center}
	\includegraphics[width=\figlenB]{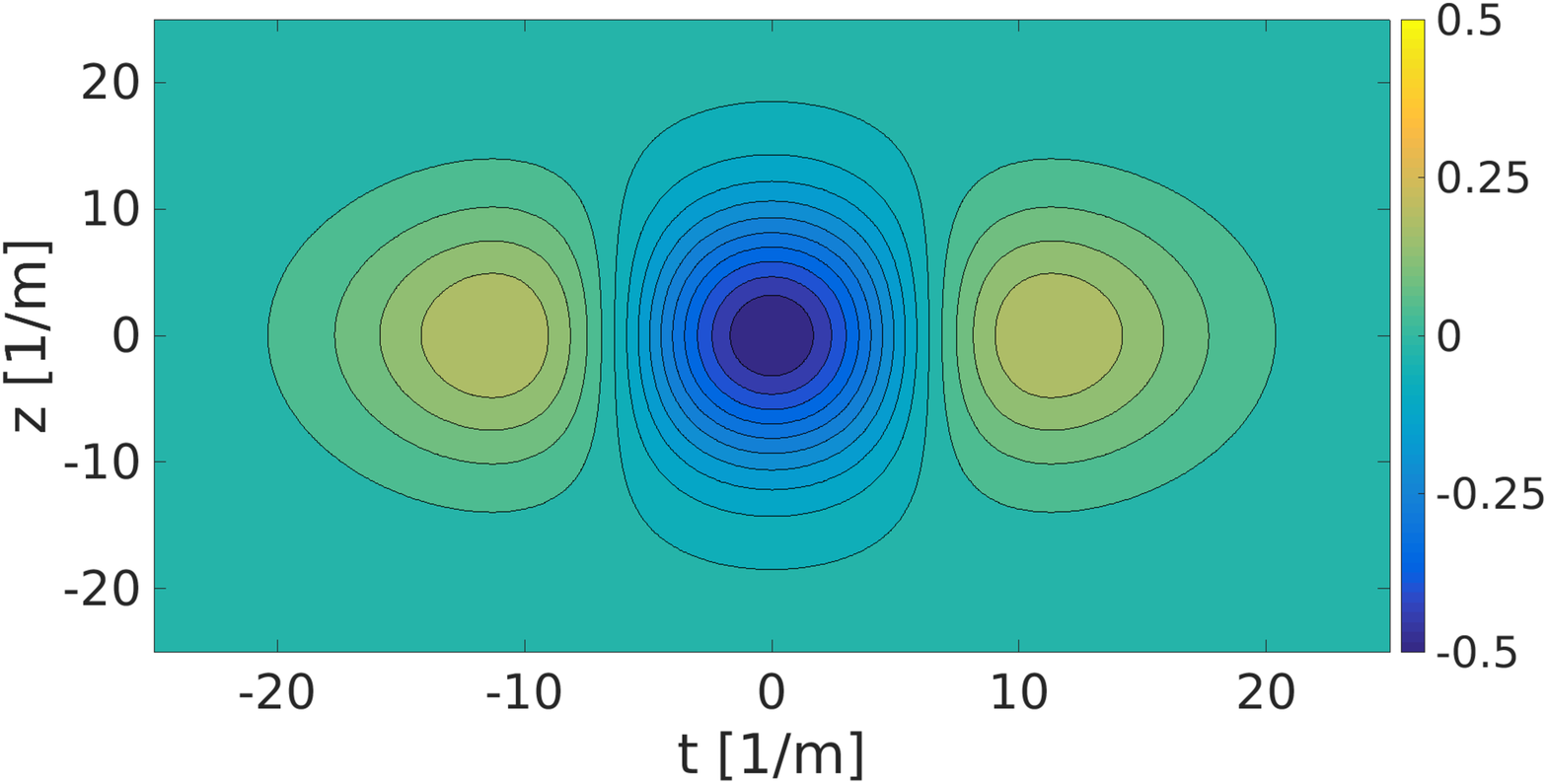} 
	\includegraphics[width=\figlenB]{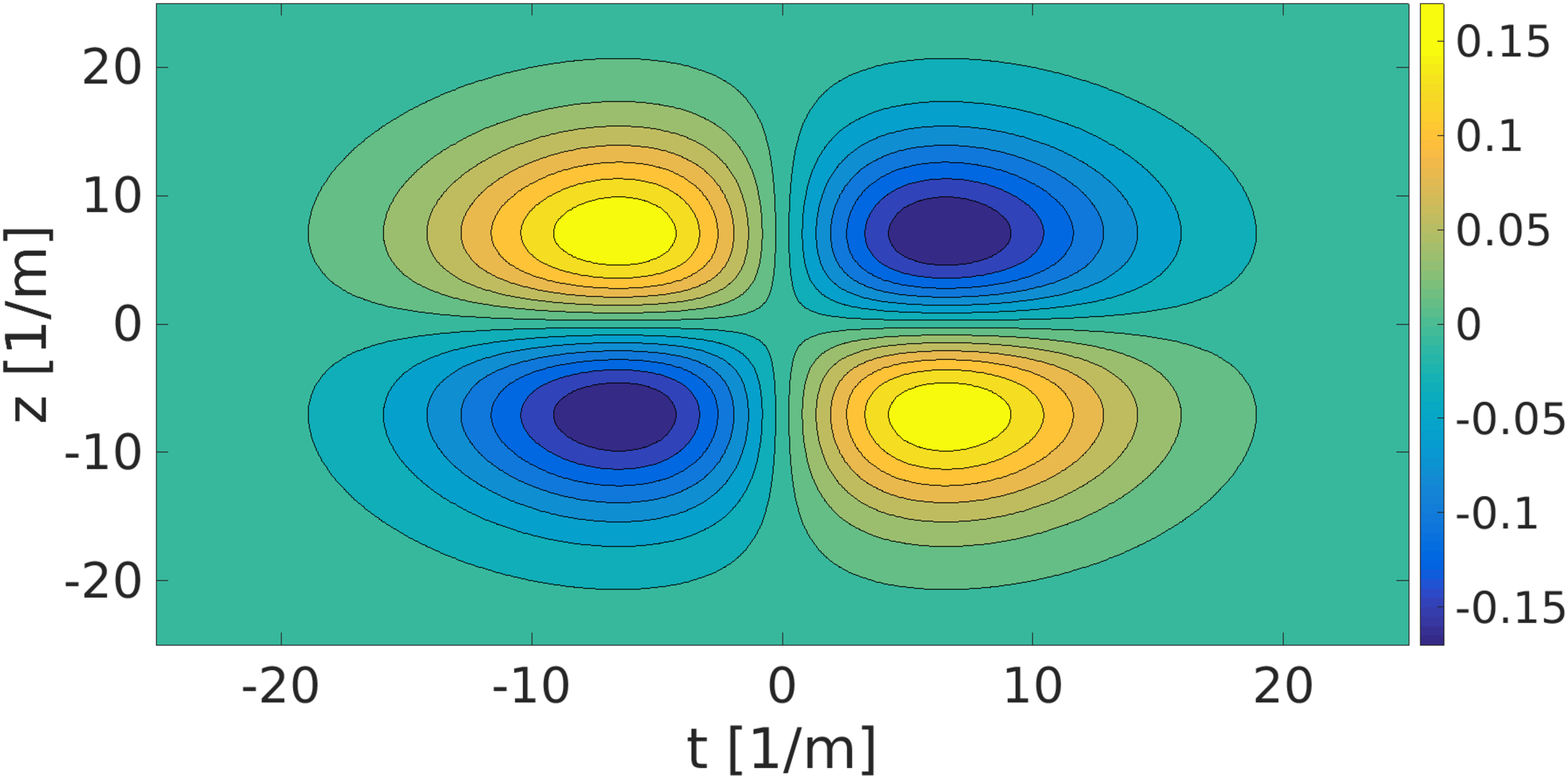} 	
      \end{center}
      \caption{Density plot of an electric field \txt{(left)}{(top)} and its corresponding magnetic field \txt{(right)}{(bottom)} as functions of space $z$ and time $t$ for $\phi=0$.
	This configuration features a single prominent peak in the electric field. Particles created at the main peak are exposed to steep field gradients of the magnetic field.
	Further parameters: $\varepsilon = 0.5$, $\tau = 20$ $m^{-1}$, $\omega = 0.1$ $m$ and $\lambda = 10$ $m^{-1}$.}
      \label{fig:EB1}
      \end{figure}

\begin{figure}[t]
      \begin{center}
	\includegraphics[width=\figlenB]{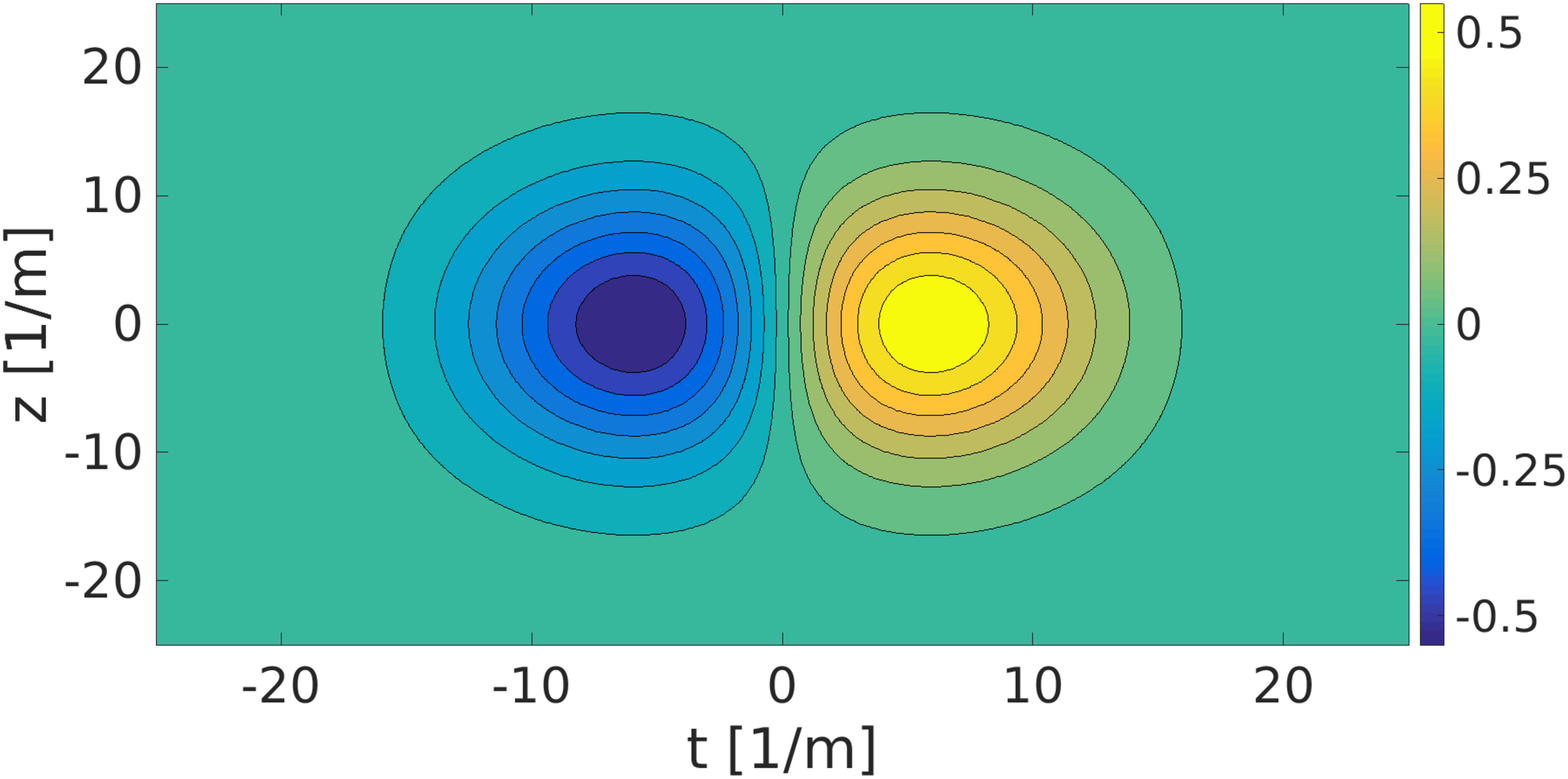} 
	\includegraphics[width=\figlenB]{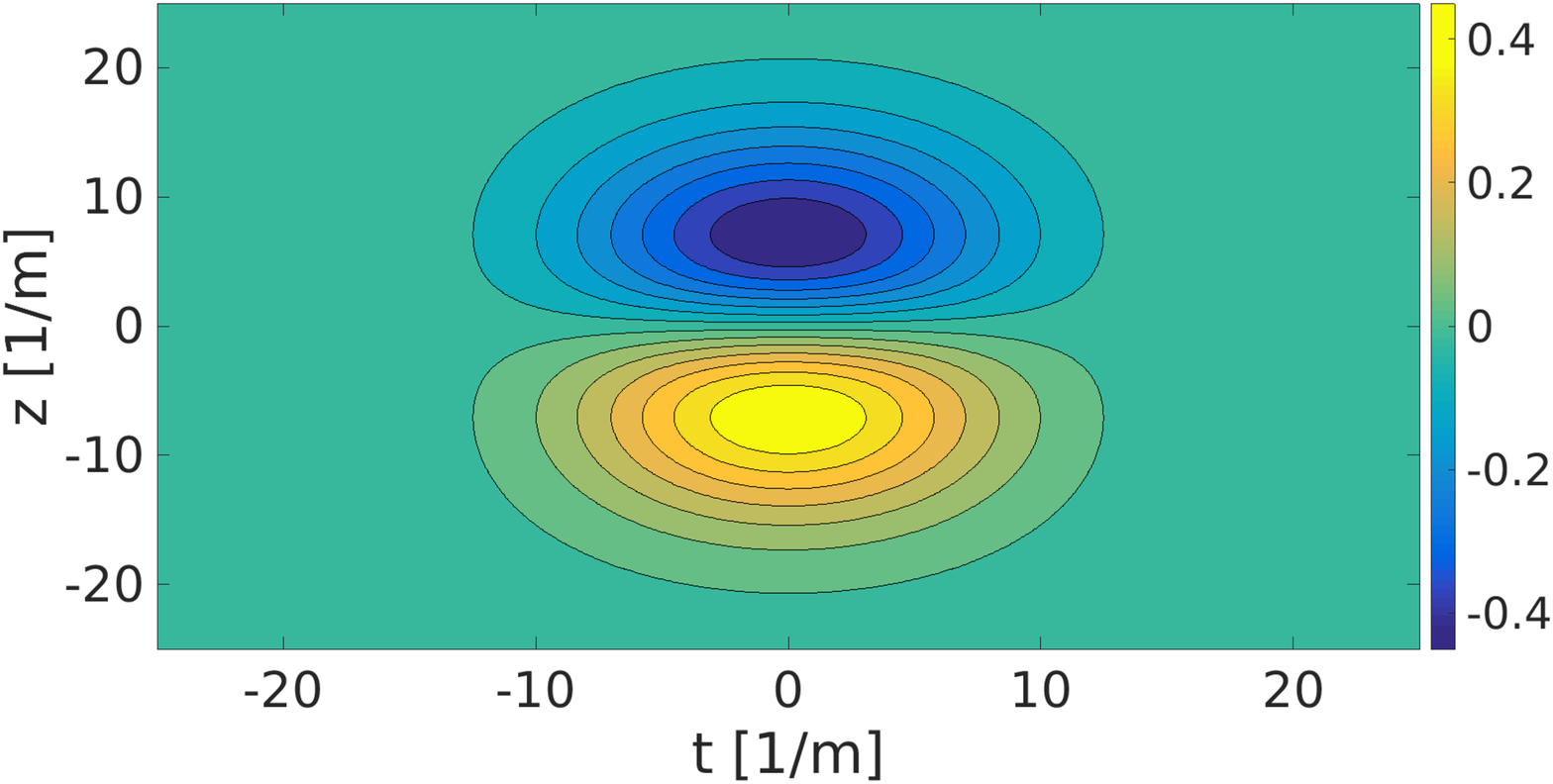} 		
      \end{center}
      \caption{Density plot of an electric field \txt{(left)}{(top)} and its corresponding magnetic field \txt{(right)}{(bottom)} as functions of space $z$ and time $t$ for $\phi=\pi/2$.
	This configuration features two peaks of equal strength in the electric field. The magnetic field is strongest between these peaks.
	Further parameters: $\varepsilon = 0.5$, $\tau = 20$ $m^{-1}$, $\omega = 0.1$ $m$ and $\lambda = 10$ $m^{-1}$.}
      \label{fig:EB2}
      \end{figure}

In this way the homogeneous Maxwell equations are automatically fulfilled and
both fields fall off sufficiently fast at asymptotic times, see Fig.~\ref{fig:EB1} and Fig.~\ref{fig:EB2}.
Moreover, it is convenient to work with fields of the form of Eqs. \eqref{equ:E} and \eqref{equ:B}, because \ck{(i) $\boldsymbol{\nabla} \cdot \mathbf{E} (t,z) = 0$ and (ii)}
the relevant phase-space can be drastically reduced. 

\section{Phase-space formalism}

The main advantage of the DHW formalism is its generality as it incorporates all features
of quantum field theory. The downside is that only few analytical results are known up to date, thus one has to implement a numerical scheme
in order to solve the governing equations \cite{Berenyi:2013eia,Kohlfurst:2015zxi,Kohlfurst:2015niu,Hebenstreit:2011wk}. This, however, turns out to be only feasible for selected field configurations due to limitations in
available computer power. 

\subsection{Theoretical foundations}
\label{ch:Theo}

The DHW formalism has been developed in Refs. \cite{Vasak:1987umA,Vasak:1987umB}. Moreover, in Refs. \cite{BB,Kohlfurst:2015zxi} one can find additional information covering all essential features of quantum kinetic approaches.
Nevertheless, we want to allow for a gentle introduction into this work. Hence, we define the covariant Wigner operator $\hat{\mathcal W}_{\alpha \beta}$ and proceed going through all the important steps, 
that are necessary to obtain the equations of motion in the end. 
Throughout this paper we use natural units $\hbar=c=1$ and express all quantities in terms
of the electron mass. In the following, we use $m_e = m$ for the mass term. 

In principle, a phase-space approach can take into account all quantum effects on the basis of the full QED Lagrangian
\eq$
\mathcal L \left( \Psi, \bar{\Psi}, A \right) = \frac{1}{2} \left( \I \bar{\Psi} \gamma^{\mu} \mathcal{D}_{\mu} \Psi - \I \bar{\Psi} \mathcal{D}_{\mu}^{\dag} \gamma^{\mu} \Psi \right) \br
 -m \bar{\Psi} \Psi - \frac{1}{4} F_{\mu \nu} F^{\mu \nu}, \label{equ:Lag}
$
where $\mathcal{D}_{\mu} = \left( \partial_{\mu} +\I e A_{\mu} \right)$ and $\mathcal{D}_{\mu}^{\dag} = \left( \overset{\leftharpoonup} {\partial_{\mu}} -\I e A_{\mu} \right)$. We have
introduced the vector potential $A_{\mu}$, the electromagnetic field strength tensor $F_{\mu \nu} = \partial_{\mu} A_{\nu} - \partial_{\nu} A_{\mu}$ 
and the spinor fields $\Psi$ and $\bar{\Psi}$.

The foundation of this phase-space approach is the density operator
\begin{equation}
 \hat {\mathcal C}_{\alpha \beta} \left( r , s \right) = \mathcal U \left(A,r,s 
\right) \ \left[ \bar {\Psi}_\beta \left( r - s/2 \right), {\Psi}_\alpha \left( r + 
s/2 \right) \right],
\end{equation}
supported by the Wilson line factor ensuring gauge invariance
\begin{equation}
 \mathcal U \left(A,r,s \right) = \exp \left( \mathrm{ie} \int_{-1/2}^{1/2} d 
\xi \ A \left(r+ \xi s \right) \ s \right).
\end{equation}
Here, we introduced the center-of-mass coordinate $r$ and the relative coordinate $s$.
A Fourier transform with respect to the relative coordinate $s$ then yields the covariant Wigner operator
\begin{equation}
 \hat{\mathcal W}_{\alpha \beta} \left( r , p \right) = \frac{1}{2} \int d^4 s \ 
\mathrm{e}^{\mathrm{i} ps} \  \hat{\mathcal C}_{\alpha \beta} \left( r , s 
\right). \label{equ:W}
\end{equation}

In order to derive the equations of motion in the DHW formalism we combine the Dirac equation
\begin{align}
  \left(\I \gamma^{\mu} \partial_{\mu} - e \gamma^{\mu} A_{\mu} - m \right) \Psi &= 0, \\
  \bar{\Psi} \left(\I \overset{\leftharpoonup} {\partial_{\mu}} \gamma^{\mu} + e \gamma^{\mu} A_{\mu} + m \right) &= 0,
\end{align}
with derivatives of the Wigner operator \eqref{equ:W}. In turn, we obtain two coupled operator equations
\begin{alignat}{3}
 & \left( \frac{1}{2} D_{\mu}  - \I P_{\mu}  \right) \gamma^{\mu} \hat{\mathcal W} \left( r , p \right) && = - &&\I \hat{\mathcal W} \left( r , p \right), \label{equ:W1} \\
 & \left( \frac{1}{2} D_{\mu}  + \I P_{\mu}  \right) \hat{\mathcal W} \left( r , p \right) \gamma^{\mu}  && = &&\I \hat{\mathcal W} \left( r , p \right), \label{equ:W2} 
\end{alignat}
with the pseudo-differential operators
\begin{alignat}{4}
 & D_{\mu}  && = \partial_{\mu}^r - e &&\int_{-1/2}^{1/2} d \xi \ && F_{\mu \nu} \left( r - \I \xi \partial^p \right) \partial_p^{\nu}, \\
 & P_{\mu}  && = p_{\mu} - \I e && \int_{-1/2}^{1/2} d \xi \ \xi \ && F_{\mu \nu} \left( r - \I \xi \partial^p \right) \partial_p^{\nu}.
\end{alignat}
Before we proceed by taking the vacuum expectation value of Eqs. \eqref{equ:W1} and \eqref{equ:W2}, we implement a simplification of Hartree type (mean-field)
\begin{equation}
 \langle \Phi | \hat F^{\mu \nu} \left( r \right) | \Phi \rangle \approx F^{\mu \nu} \left( r \right)
\end{equation}
transforming the operator-valued electromagnetic field strength tensor to a C-number field. 
Hence, terms of the form $\hat F^{\mu \nu} \hat{\mathcal C}$ simply become
\begin{equation}
 \langle \Phi | \hat F^{\mu \nu} \left( r \right) \ \hat{\mathcal C} \left( r , s \right) | \Phi \rangle 
  \approx F^{\mu \nu} \left( r \right) \langle \Phi | \hat{\mathcal C} \left( r , s \right) | \Phi \rangle.
\end{equation}
As a result, we obtain an equation of motion for the covariant Wigner function
\begin{equation}
 \mathcal W \left( r , p \right) = \langle \Phi | \hat{\mathcal W} \left( r , p \right) | \Phi \rangle
\end{equation}
and subsequently, as we are interested in a time-evolution formalism, for the equal-time Wigner function
\begin{equation}
 {\mathbbm w} \left( t, \mathbf{x} , \mathbf{p} \right) = \int \frac{dp_0}{2 \pi} \mathcal W \left( r , p \right).
\end{equation}


In order to make calculations in inhomogeneous fields feasible we have to reduce the available phase-space. \ck{Due to the special form of the background fields, Eqs. \eqref{equ:A}, \eqref{equ:E}
and \eqref{equ:B}, particle dynamics can be confined to the $xz$-plane, (i) via using a separation ansatz for the $3$-dimensional formalism (by fixing $p_y=0$) or (ii) 
by deriving the equations of motions using a $2d$ QED Lagrangian as basis, cf. Ref. \cite{Kohlfurst:2015niu,Kohlfurst:2015zxi} for detailed derivations.  
Either way, the available particle phase-space is greatly reduced, which, in turn, leads to a reduced Wigner function $\overline{\mathbbm{w}} \left( t, \mathbf x, \mathbf p \right)$.}
Decomposing it into Dirac bilinears using a $2$-spinor formulation yields\footnote{The non-consecutive indices are chosen to put emphasis on the idea of
working in a subspace of the whole phase-space ($p_y=0$), see Refs. \cite{Hebenstreit:2011wk,Kohlfurst:2015niu,Kohlfurst:2015zxi} for comparison.}
\begin{equation}
 \overline{\mathbbm{w}} \left( t, \mathbf x, \mathbf p \right) = \frac{1}{2} \left( \mathbbm 1 \ \mathbbm s + \gamma_{0} \mathbbm v^{0} + \gamma_{1} \mathbbm v^{1} + \gamma_{3} \mathbbm v^{3} \right). \label{equ:wigner}
\end{equation}
Following Ref. \cite{BB} we can interpret $\mathbbm s$ as mass density, $\mathbbm{v}_0$ as charge density and 
$\mathbbm v^{1},~ \mathbbm v^{3}$ as a current density vector. \ck{It is worth mentioning, that as a side effect of a $2d$ formulation, 
the magnetic field as well as the particle spin and the particles angular momentum become scalar quantities \cite{Kohlfurst:2015zxi}. 
Furthermore, the underlying $2$-spinor formulation inherently favors one spin direction.
\footnote{The definition above represents only one of various possible reductions corresponding to a $2$-spinor derivation of the Wigner function using a $2d$ QED
Lagrangian as basis, cf. Ref. \cite{Kohlfurst:2015niu,Kohlfurst:2015zxi}. }
} 

Eventually, we obtain a coupled set of equations of motions for the Wigner coefficients, cf. Refs. \cite{Kohlfurst:2015niu,Kohlfurst:2015zxi},
\begin{alignat}{6}
  & D_t \mathbbm{v}_0 && + D_x \mathbbm{v}^1 && + D_z \mathbbm{v}^3 && && &&= 0, 
\label{eqn1_1} \\  
  & D_t \mathbbm{s}  && && && -2 \Pi_x \mathbbm{v}^3 && +
 2 \Pi_z \mathbbm{v}^1 &&= 0,  \label{eqn1_2} \\  
  & D_t \mathbbm{v}^1 && +D_x \mathbbm{v}_0 &&  && && -2 \Pi_z
 \mathbbm{s} && = -2 \mathbbm{v}^3,  \label{eqn1_3} \\  
  & D_t \mathbbm{v}^3 && && +D_z \mathbbm{v}_0 && +2\Pi_x
 \mathbbm{s} && &&= 2 \mathbbm{v}^1, \hspace{0.7cm}  \label{eqn1_4} 
\end{alignat} 
with the pseudo-differential operators
\begin{alignat}{8}
  & D_t && = \quad && \partial_{t} && + e && \int_{-1/2}^{1/2} d\xi \, && \mathbf{E} \left(
\mathbf{x}+ \textrm {i} \xi \boldsymbol{\nabla}_p,t \right) && \cdot && \boldsymbol{\nabla}_p,
\label{eqn2_1} \\
 & D_x && = \quad && \partial_x && + e && \int_{-1/2}^{1/2} d\xi \, && B \left( \mathbf{x}+\textrm {i} \xi \boldsymbol{\nabla}_p,t \right) && && \partial_{p_z},
\label{eqn2_2} 
\end{alignat}
\begin{alignat}{8}
 & D_z && = \quad && \partial_z && - e && \int_{-1/2}^{1/2} d\xi \, && B \left( \mathbf{x}+\textrm {i} \xi \boldsymbol{\nabla}_p,t \right) && && \partial_{p_x},
\label{eqn2_3} \\
  & \Pi_x && = \quad && p_x && - \textrm {i} e && \int_{-1/2}^{1/2} d\xi \,
\xi \, && B \left( \mathbf{x}+\textrm {i} \xi \boldsymbol{\nabla}_p,t \right) && && \partial_{p_z}, \label{eqn2_4} \\
  & \Pi_z && = \quad && p_z && + \textrm {i} e && \int_{-1/2}^{1/2} d\xi \,
\xi \, && B \left( \mathbf{x}+\textrm {i} \xi \boldsymbol{\nabla}_p,t \right) && && \partial_{p_x}. \label{eqn2_5} 
\end{alignat}
As we have restricted the available phase-space volume to the $xz$-plane, spatial and momentum vectors are given via
\begin{equation}
 \mathbf{x} = \left( x , z \right)^T, \qquad \mathbf{p} = \left( p_x , p_z \right)^T.
\end{equation}

In the following we incorporate vacuum initial conditions
\begin{alignat}{3}
  \mathbbm{s}_{vac} \left(\boldsymbol{p} \right) = -\frac{2}{\sqrt{1 +
\boldsymbol{p}^2}}, \quad && 
  \mathbbm{v}_{vac}^{1,3} \left(\boldsymbol{p} \right) = -\frac{2
\boldsymbol{p}}{\sqrt{1 + \boldsymbol{p}^2}} \label{equ:vac}
\end{alignat}
into the system of equations \eqref{eqn1_1}-\eqref{eqn1_4} by switching to modified Wigner components \cite{Hebenstreit:2011wk}
\begin{equation}
 \overline{\mathbbm{w}}^v = \overline{\mathbbm{w}} - \overline{\mathbbm{w}}_{vac}.
 \label{equ:red}
\end{equation}
In this way, Eqs. \eqref{eqn1_1}-\eqref{eqn1_4} are turned
into a set of inhomogeneous partial differential equations. The particle number density 
\begin{align}
n \left( z, p_x, p_z \right) = \frac{\mathbbm{s}^v + p_x
\mathbbm{v}^{v,1} + p_z \mathbbm{v}^{v,3}}{\sqrt{1+\boldsymbol{p}^2}}
 \label{equ:n}
\end{align}
as well as the charge density 
\begin{align}
c \left( z, p_x, p_z \right) = e \, \mathbbm{v}_0^v
 \label{equ:c}
\end{align}
are defined for asymptotic times $t_f$ ($\mathbf A (t_f,z) \to 0$). 

Furthermore, the particle's momentum distribution per unit volume 
\begin{align}
 &n \left( p_x, p_z \right) &&= \int dz ~ n \left( z, p_x ,p_z \right), \\
 &n \left( p_x \right) &&= \int dz ~ dp_z ~ n \left( z, p_x ,p_z \right), \\ 
 &n \left( p_z \right) &&= \int dz ~ dp_x ~ n \left( z, p_x ,p_z \right) \hspace{2cm}
 \label{equ:nn}
\end{align}
as well as the particle's position-momentum distribution per unit volume 
\begin{align}
 &n \left( z, p_z \right) = \int d p_x ~ n \left( z, p_x ,p_z \right), 
\end{align}
are derived from Eq.~\eqref{equ:n}.

\subsection{Solution strategy}
\label{ch:Sol}

The equations of motion \eqref{eqn1_1}-\eqref{eqn1_4} are solved numerically in the vicinity of background fields given by Eqs. \eqref{equ:E} and \eqref{equ:B}.
As the fields are homogeneous in $x$, the domain is three-dimensional.
No further truncation is applied, see Refs. \cite{Kohlfurst:2015niu,Kohlfurst:2015zxi} for alternative strategies.
Nevertheless, we enhance numerical stability at reduced computational costs by introducing a transformation of variables \cite{Kohlfurst:2015zxi} of the form
\begin{align}
 p_x &= \frac{2 L_q}{\pi } \textrm{arctan} \left( \frac{1}{\alpha_{q}} \, \textrm{tan} \left( \frac{\pi}{2 L_q} \, q_x \right) \right), \\
 z &= \frac{2 L_z}{\pi } \textrm{arctan} \left( \frac{1}{\alpha_{z}} \, \textrm{tan} \left( \frac{\pi}{2 L_z} \, \tilde z \right) \right).
\end{align}
The quantities $L_q$ and $L_z$ give the length in $p_x$- and $z$-direction, respectively. The parameters $\alpha_{q}$ and $\alpha_{z}$ control the strength of the
deformation, with $\alpha=1$ corresponding to the identity transformation.
In turn, the differential operators are transformed accordingly
\begin{align}
 \partial_{p_x} &= \left( \alpha_{q} \ \textrm{cos} \left( \frac{\pi}{2 L_q} \, q_x \right)^2 + \frac{1}{\alpha_{q}} \ \textrm{sin} \left( \frac{\pi}{2 L_q} \, q_x \right)^2 \right) \partial_{q_x}, \\
 \partial_z &= \left( \alpha_{z} \ \textrm{cos} \left( \frac{\pi}{2 L_z} \, \tilde z \right)^2 + \frac{1}{\alpha_{z}} \ \textrm{sin} \left( \frac{\pi}{2 L_z} \, \tilde z \right)^2 \right) \partial_{\tilde z}. 
\end{align}

Due to the fact, that we have already taken care of the initial conditions \eqref{equ:red} all reduced Wigner functions vanish for high $\tilde z$, $q_x$ and $p_z$.  
This allows us to artificially demand periodic boundary conditions in spatial and momentum coordinates, 
thus transforming the flat phase-space domain to a toroidal domain.

The transformed system of PDEs is then solved by taking advantage of the method of lines. More precisely, the domain in phase-space is equidistantly discretized
leaving the time variable $t$ as the only continuous parameter. To account for the boundary conditions we simply set $\tilde z_0 = \tilde z_{N_z}$, where
$N_z$ is the total number of grid points in $\tilde z$. The same procedure is applied to the variables $q_x$ and $p_z$.
This discretization allows to solve the differential equation with spectral methods on a Fourier basis at given time $t$ \cite{Boyd,Tref}. 

The general procedure to calculate derivatives is then given by
\begin{equation}
 \mathcal{FT}^{-1} \left[ \mathcal{FT} \left[ \frac{d^n}{dx^n} f \left( x \right) \right] \right] = \mathcal{FT}^{-1} \left[ \left( \I \kk \right)^n \hat f \left( \kk \right) \right],
\end{equation}
where $\hat f \left( \kk \right)$ denotes a Fourier transformed quantity.
Evaluating the pseudo-spectral differential operators \eqref{eqn2_1}-\eqref{eqn2_5} is more involved due to the appearance of derivatives as arguments of functions.
That is, when the essential advantage of the pseudo-spectral method comes into play. Exemplary for any non-local terms in Eqs. \eqref{eqn2_1}-\eqref{eqn2_5}, we write
\begin{equation}
 \Delta \mathbbm{w}^v \left( \tilde z, q_x, p_z \right) = e \int d\xi \, G \left( \tilde z + \I \xi \partial_{p_z},t \right) \ \mathbbm{w}^v \left( \tilde z, q_x, p_z \right)
\end{equation}
for a generic modified Wigner component. Applying pseudo-spectral methods we obtain
\eq$
 \mathcal{FT}^{-1} \ \left[ \mathcal{FT} \ \left[ \Delta \mathbbm{w}^v \left( \tilde z, q_x, p_z \right) \right] \right] = \br
 \mathcal{FT}^{-1} \ \left[ e \int d\xi \, G \left( \tilde z - \xi \kk_{p_z} ,t \right) \hat{\mathbbm{w}}^v \left( \tilde z, q_x, \kk_{p_z} \right) \right].
$
Due to the special form of the vector potential \eqref{equ:A}, (i) the time dependency can be factored out and (ii) the integral can be performed analytically, eventually leading to
\eq$
 \Delta \mathbbm{w}^v \left( \tilde z, q_x, p_z \right) = \br
 \overline G \left( t \right) \ \mathcal{FT}^{-1} \ \left[ \tilde G \left( \tilde z, \kk_{p_z} \right) \hat{\mathbbm{w}}^v \left( \tilde z, q_x, \kk_{p_z} \right) \right].
$
Hence, we have successfully transformed a non-local differential operator into a simple multiplicative factor. Computational costs can then be
reduced further by applying anti-aliasing procedures, e.g. termination of the highest wave numbers \cite{Boyd}. 
Additionally, the inhomogeneous source terms do not need to be solved spectrally. A Taylor expansion in the momentum variables (up to eighth order)
turned out to be completely sufficient for the fields under investigation in this article.
In order to perform the time integration, we rely on a Dormand-Prince Runge-Kutta integrator
of order 8(5,3) \cite{NR}.\footnote{Technical aspects: Computations were performed on
Supermicro Servers. The calculations were done in parallel cumulating in a total CPU time of
$10$ $d$ ($\tau=10$ $m^{-1}$) and $70$ $d$ ($\tau=20$ $m^{-1}$), respectively.
The grid size in phase-space was $512\times512\times512$ ($\tau=10$ $m^{-1}$) and $768\times512\times512$ ($\tau=20$ $m^{-1}$), respectively.}

\subsection{Coarse graining}
\label{ch:CG}

The way the Wigner function is defined, an interpretation in terms of real observables is technically only allowed if they are given either in momentum or in spatial coordinates \cite{moyal_1949A,moyal_1949B}. 
In case of mixed representations, the Wigner method yields only quasi-probabilities which, in turn, makes discussions generally vague. \ck{To overcome this systematic
handicap, we have implemented a coarse graining technique assuming that the unphysical parts vary more rapidly than the physical quantities. 
For comparison, it was already shown in Ref. \cite{Rafelski:1993uh}, that the Wigner method induces highly oscillating terms; $\cos \bigg(x \ \Big(p_1-p_2 \Big) \bigg)$.
As we investigate Schwinger pair production for spatial variations of the order of \mbox{$\lambda \gg 1$}, we therefore expect averaging techniques to hold reasonably 
well.}

Coarse graining methods are an important tool in order to study, e.g., chemical processes \cite{ChemieA,ChemieB}. Proper application of coarse-grained modeling
significantly reduces the number of degrees of freedom, while the relevant information is retained. This enabled the study of time-evolutions 
of large complex structures, cf. polymer melting \cite{Poly} or molecular dynamics \cite{BioMolA,BioMolB,BioMolC}. 

A related coarse graining technique has already been introduced in Ref. \cite{Rafelski:1993uh} to study the relativistic classical limit of the DHW formalism.
In this work, a Gaussian-type smearing function was introduced 
\eq$
 G \left( \mathbf{x} - \mathbf{x}', \mathbf{p}-\mathbf{p}' \right) = \br
 \left( \frac{1}{\pi \lambda_x \lambda_p} \right)^3 
  \ee^{-\left(\mathbf{x}-\mathbf{x}' \right)^2/ \lambda_x^2 -\left(\mathbf{p}-\mathbf{p}' \right)^2/ \lambda_p^2 }, \label{equ:Fold}
$
with the coarseness parameters $\lambda_x \lambda_p \gg 1$. The convolution of the Wigner function \eqref{equ:wigner} with Eq.~\eqref{equ:Fold} then yields 
coarse grained versions of the phase-space functions. Moreover, it could be shown, that these functions give the correct classical limit \cite{Rafelski:1993uh}.
We expand the idea to obtain meaningful results also in the quantum regime. In contrast to the previous work,
we apply the smearing function only at asymptotic times. In this way, we do not introduce further truncations, while simultaneously
eliminating the constraint on working in either a spatial or a momentum representation.

Hence, we implement the coarse graining as a post-processing step within the phase-space approach. As we are mainly interested in particle distributions in
$z p_z$-space, we first calculate $n \left(z , p_z \right)$ and then apply the discrete smearing operator
\begin{equation}
 G_{mn} \left( z_i, p_{z,j} \right) = \frac{1}{Z_G} \ \ee^{-\frac{ \left( z_m - z_i \right)^2}{2 \sigma_z^2} -\frac{ \left( p_{z,n} - p_{z,j} \right)^2}{2 \sigma_{p_z}^2} }
 \label{equ:Gmn}
\end{equation}
for every phase-space coordinate $\left( z_i, p_{z,j} \right)$. Indices run over $m = i-M_{\rm z}, \ldots, i+M_{\rm z}$ and $n = j-M_{\rm p_z}, \ldots, j+M_{\rm p_z}$, where
$M_z$ and $M_{p_z}$ determine the domain.
Additionally, the operator $G_{mn} \left( z_i, p_{z,j} \right)$ is normalized to unity preserving the total yield. 
For the sake of simplicity, we assumed a quadratic stencil leading to $M = M_{\rm z} = M_{\rm p_z}$ and $\sigma = \sigma_z = \sigma_{p_z}$. 

\section{Single-particle trajectory analysis}
\label{ch:Traj}

An alternative approach towards studying momentum resolved particle production is given via analysis of the trajectories of ``randomly'' 
created particle pairs. Here, we present a semi-classical model combining effective field theory with
classical equations of motion.

Instead of dealing with the full QED Lagrangian \eqref{equ:Lag} we introduce the Heisenberg-Euler Lagrangian \cite{Heisenberg:1935qt} at this point
\eq$
 {\cal L}_{EH} \left(a, b \right) = -\frac{1}{8 \pi^2} \int_0^\infty \, \frac{d \eta}{\eta^3} \, \ee^{-\eta e E_{cr}} \br
  \times \left( \frac{e^2 a b \eta^2}{{\rm tanh} \left(eb \eta \right) {\rm tan} \left(ea \eta \right)} -1 - \frac{e^2 \eta^2}{3} \left(b^2 - a^2 \right) \right). \label{equ:LEH}
$
The quantities $a$ and $b$ play a decisive role as they are connected to the Lorentz invariants
\begin{align}
 a^2 - b^2 = \mathbf{E}^2 - \mathbf{B}^2 = -\frac{1}{2} F_{\mu \nu} F^{\mu \nu} = -2 {\cal F}, \\
 a \, b = \mathbf{E} \cdot \mathbf{B} = -\frac{1}{4} F_{\mu \nu} \tilde F^{\mu \nu} = - {\cal G}.
\end{align}
Hence, they can be expressed as
\begin{equation}
 a = \sqrt{ \sqrt{ {\cal F}^2 + {\cal G}^2} - {\cal F}}, \qquad b = \sqrt{ \sqrt{ {\cal F}^2 + {\cal G}^2} + {\cal F}}.
\end{equation}

Analysis of Eq.~\eqref{equ:LEH} for constant perpendicular fields, ${\cal G} = -\mathbf{E} \cdot \mathbf{B} = 0$, reveals three different possibilities \cite{Dunne:2004nc}:
If ${\cal F} > 0$ then $\mathbf{B}^2 > \mathbf{E}^2$ and pair production is not possible. If ${\cal F}$ vanishes, then there are no quantum corrections at all.
Only in case of ${\cal F} < 0$ the formation of particles is allowed. 

As we are interested in pair production we concentrate on the case $\mathbf{E}^2 > \mathbf{B}^2$ and $\mathbf{E} \perp \mathbf{B}$. This automatically implies, that $b=0$ and 
$a=\sqrt{ \lvert {\cal F} \rvert - {\cal F} }$. \footnote{There has been a notational revision compared to Refs. \cite{Kohlfurst:2015niu,Kohlfurst:2015zxi}. 
Nevertheless, the ``effective field amplitude'', 
that has been introduced in these works, yields the same information.} In turn, we obtain
\eq$
 {\cal L}_{EH} \left(a, 0 \right) = \br 
  -\frac{1}{8 \pi^2} \int_0^\infty \, \frac{d \eta}{\eta^3} \, \ee^{-\eta e E_{cr}} 
  \left( \frac{e a \eta}{{\rm tan} \left(ea \eta \right)} -1 + \frac{e^2 \eta^2}{3} a^2 \right). \label{equ:LEH2}
$
\ck{Analyzing Eq. \eqref{equ:LEH2}, we see that Schwinger pair production is exponentially suppressed, where 
$eE_{cr} = m^2$ yields the limit at which the work an electric field does on a particle pair over the Compton wave length equals the pairs rest energy.
Additionally, in case of linearly polarized fields the threshold for particle production can be altered if initial transversal momenta $p_T$ are taken into account, see Ref. \cite{Casher:1979gwA,Casher:1979gwB}.
In such a case the particles initial energy $E_T = \sqrt{m^2 + p_T^2}$ and thus the field strength it takes to create particles at the same rate increases $eE_{cr} \to m^2 + p_T^2$.
}
Following Ref. \cite{Schwinger:1951nm} and assuming that $a$ is sufficiently smaller
than the critical field strength we extract an estimate for the formation 
rate\footnote{Generally, the probability for pair production to happen is given in terms of a vacuum decay rate \cite{Cohen:2008wz,Hebenstreit:2008ae,Narozhnyi,Lin:1998rn}, which does not allow for a momentum resolved investigation.} 
of single electron-positron pairs in linearly polarized fields \cite{Casher:1979gwA,Casher:1979gwB}
\begin{equation}
 \dot N \approx \frac{e a}{4 \pi^3} \exp \left(- \frac{\pi \left(m^2 + p_T^2 \right)}{e a} \right), \label{equ:N}
\end{equation}

The relations above have been derived for constant electric and magnetic fields. Nevertheless, these equations are known to hold also
for slowly varying fields with corrections governed by $\left( \frac{\partial}{m} \right)^2$, where $\partial$ denotes the variation scale of the field \cite{Galtsov,Karbstein:2017pbf}. 
Hence, having fields of the form of Eqs. \eqref{equ:E} and \eqref{equ:B} in mind, 
we will use Eq.~\eqref{equ:N} in a locally constant field approximation as a probability weight to resemble an instantaneous source term 
for particle production in an inhomogeneous background field
\begin{equation}
 P \left(t , z, p_z \right) = \frac{e a \left(t , z \right)}{4 \pi^3} \exp \left(- \frac{\pi \left(m^2 + p_z^2 \right)}{e a\left(t , z \right)} \right). \label{equ:P}
\end{equation}
To be more specific, we first create a sample of random variables $\left(t_{0i}, z_{0i}, p_{z_{0i}} \right)$. Then we test each tuple for the likeliness of pair production
at time $t_{0i}$ at coordinate $z_{0i}$ and initial transversal momentum $p_{z_{0i}}$ by comparing $P \left(t_{0i} , z_{0i}, p_{z_{0i}} \right)$ with a random variable $\rho = \textrm{rand} \left( 0, \textrm{max} \left( P \right) \right)$.
If $P \left(t_{0i} , z_{0i}, p_{z_{0i}} \right) > \rho$ the coordinates are accepted for further treatment. 

After the particles have been created they are deflected by the electromagnetic background field. In the simplest approximation, the electrons/positrons are assumed to follow
``classical'' trajectories, where, e.g., radiation effects can be ignored. Moreover, we want to assume that these particles do not interact with each other, thus allowing to
inspect their trajectories one-by-one. At this point, we give up quantum mechanical phase information. Hence, we trade the ability to describe e.g. quantum interferences for an easier numerical implementation.
However, this approach still helps to properly
set up the parameters for a full DHW calculation.

The actual calculation of the particle trajectories is done via a modified relativistic Lorentz force equation \cite{Silenko:2007wi,MengWenA,MengWenB}
\begin{equation}
 \frac{d u^\alpha}{d \tau} = e F^{\alpha \beta} u_\beta + f_s^\alpha, \label{equ:L}
\end{equation}
where $u^\alpha$ is the four-velocity, $F^{\alpha \beta}$ the electromagnetic field strength tensor and $f_s^\alpha$ is an additional model-specific, spin-dependent force.
Based upon the analysis provided in Ref. \cite{MengWenA,MengWenB} we decided to choose a Foldy-Wouthuysen-like model \cite{Silenko:2007wi}.
Omitting the anomalous magnetic moment of the electron we define \cite{Silenko:2007wi}: 
\begin{equation}
 \mathcal H = \gamma^0 \left( 1 + \boldsymbol{\gamma} \cdot \hat{\mathbf{p}} \right),
\end{equation}
with the kinetic momentum operator $\hat{\mathbf{p}}$. Performing the Foldy-Wouthyusen transformation yields
\begin{multline}
 \mathcal H_{FW} = \gamma^0 \epsilon - \frac{e}{4} \left\{ \frac{1}{\epsilon}, \boldsymbol{\Pi} \cdot \mathbf{B} \right\} + \\
  \frac{e}{2 \sqrt{2 \epsilon \left(\epsilon + 1 \right)}} \bigg( \boldsymbol{\Sigma} \cdot \Big( \hat{\mathbf{p}} \times \mathbf{E} - \mathbf{E} \times \hat{\mathbf{p}} \Big) \bigg) \frac{1}{\sqrt{2 \epsilon \left(\epsilon + 1 \right)}} + 
 \mathcal{O} \left( \hbar^2 \right), \label{equ:HFW}
\end{multline}
with the spin operator $\boldsymbol{\Sigma}$, the polarization operator $\boldsymbol{\Pi} = \gamma^0 \boldsymbol{\Sigma}$ and $\epsilon = \sqrt{1+ \hat{\mathbf{p}}^2}$.
The time evolution equation for the momentum operator is given by
\begin{equation}
 \frac{d \hat{\mathbf{p}} }{dt} = \textrm{i} \left[ \mathcal H_{FW}, \hat{\mathbf{p}} \right] + e \mathbf{E}.
\end{equation}
Plugging in Eq.~\eqref{equ:HFW} then yields
\begin{multline}
  \frac{d \hat{\mathbf{p}} }{dt} = \\
  e \mathbf{E} + \frac{e \gamma^0}{4} \left\{ \frac{1}{\epsilon}, \hat{\mathbf{p}} \times \mathbf{B} - \mathbf{B} \times \hat{\mathbf{p}} \right\} 
  +\frac{e}{4} \left\{ \frac{1}{\epsilon}, \boldsymbol{\nabla} \left( \boldsymbol{\Pi} \cdot \mathbf{B} \right) \right\} \\
    -\frac{e}{2 \sqrt{2 \epsilon \left(\epsilon + 1\right)}} \bigg( \boldsymbol{\nabla} \Big( \boldsymbol{\Sigma} \cdot \left( \hat{\mathbf{p}} \times \mathbf{E} -\mathbf{E} \times \hat{\mathbf{p}} \right) \Big) \bigg) \frac{1}{\sqrt{2 \epsilon \left(\epsilon + 1 \right)}}.
\end{multline}
In the semi-classical limit the operators are transformed to classical quantities. Accordingly, the evolution equation for the momentum operator becomes
an equation of motion for a particle in an external electromagnetic background field
\begin{multline}
 \frac{d \mathbf{p} }{dt} = \mathbf{F}_E + \mathbf{F}_B + \mathbf{F}_S = \\ 
  e \left( \mathbf{E} + \frac{\mathbf{p} \times \mathbf{B}}{\gamma \left(t \right)} + 
  \frac{1}{\gamma} \boldsymbol{\nabla} \left( \mathbf{s} \cdot \left( \mathbf{B} - \frac{\mathbf{p} \times \mathbf{E} }{\gamma +1} \right) \right) \right),
\end{multline}
with the classical spin vector $\mathbf{s}$ and the Lorentz factor $\gamma \left( t \right) = \sqrt{1 + \mathbf{p}^2 }$. \footnote{
In general, the particle spin precesses in an external field.
This dynamics is governed by the Thomas-Bargmann-Michel-Telegdi equation \cite{Thomas:1927yu,Bargmann:1959gz}, which can be derived in a similar fashion. 
Omitting the anomalous magnetic moment the spin motion is approximately given by \cite{Silenko:2007wi}
\begin{equation*}
 \frac{d \mathbf{s}}{dt} = \frac{e}{\gamma} \left( \mathbf{s} \times \mathbf{B} - \frac{\mathbf{s} \times \left( \mathbf{p} \times \mathbf{E} \right)}{\gamma+1} \right).
\end{equation*}
}

Due to the special form of the vector potential \eqref{equ:A} and due to the focus on lower dimensional configurations the equations can be reduced to a system of two coupled
equations. In particular the spin force term $\mathbf{F}_S$ can be greatly simplified assuming that the spin term $\mathbf{s}$ points in $y$-direction and, hence, does not precess.
Eventually, we introduce a relation between relativistic velocity and momentum to obtain a complete set of differential equations
\begin{alignat}{5}
 & \frac{\partial z}{ \partial t} &&= +\frac{p_z \left(t\right)}{\gamma \left(t\right)}, \label{equ:sin1} &&\\
 & \frac{\partial p_x}{\partial t} &&= - \frac{p_z \left(t\right)}{\gamma \left(t\right)} B \left(t,z(t) \right) + E \left(t,z(t) \right), && \label{equ:sin2} \\
 & \frac{\partial p_z}{\partial t} &&= +\frac{p_x \left(t\right)}{\gamma \left(t\right)} B \left(t,z(t) \right) && \label{equ:sin3} \\ 
  & &&+ \frac{s}{\gamma \left(t\right)} \left( \partial_z B \left(t,z(t) \right) - \frac{p_z \left(t\right)}{\gamma \left(t\right)+1} \partial_z E \left(t,z(t) \right) \right). && \notag 
\end{alignat}
The spin term reduces to $s=\pm 1/2$ and the Lorentz factor is given by $\gamma \left( t \right) = \sqrt{1 + p_x \left(t\right)^2 + p_z \left(t\right)^2 }$.
All in all, we have reduced the problem of determining the particle dynamics to solving an ordinary differential equation. The corresponding initial conditions are given by
the coordinates $(t_0, z_0, p_{z_0})$ of a particle created at $t_0$.

Solving the equations of motion \eqref{equ:sin1}-\eqref{equ:sin3} for $N$ particles gives us a total of $N$
data points in phase space. As we have sampled the initial conditions randomly, meaningful statements can only be made by evaluating the data collectively.
Hence, we use a kernel density estimation method \cite{parzen1962,Rosenblatt} in order to transfer the individual data points into a smooth density function. 
In this way, we can make
statements on the particle distributions even allowing for a comparison with the results obtained via solving the DHW equations.
As this single-particle trajectory method is not designed to study phase information it cannot describe quantum interference effects. However,
we can search the particle phase-space for overlapping data points. If, for example, two particles created at totally different times
are found close to each other in phase-space, we would expect quantum effects to play a decisive role. 

We want to emphasize, that our semi-classical model should not be confused with semi-classical methods to compute Schwinger
pair production \cite{SemiClassA,SemiClassB,SemiClassC,SemiClassD,SemiClassE,SemiClassF,SemiClassG}. The latter uses semi-classical approximations to obtain a quantum production rate in inhomogeneous fields. 
Our model, however, uses the constant field production rate locally and then extracts phase-space information from the subsequent classical trajectories. 
In fact, both approaches can in principle be combined yielding an improved estimate for the production process.

\section{Results}

In the following we present results obtained from numerically solving the DHW equations \eqref{eqn1_1}-\eqref{eqn1_4}. We interpret the particle distributions and highlight features
that are connected with the applied electric and magnetic fields. Moreover, we discuss the particle densities on the basis of the semi-classical trajectory model. 

\subsection{Coarse-grained particle distribution}
\label{ch:Coarse}

      \begin{figure}[t]
      \includegraphics[width=\figlen]{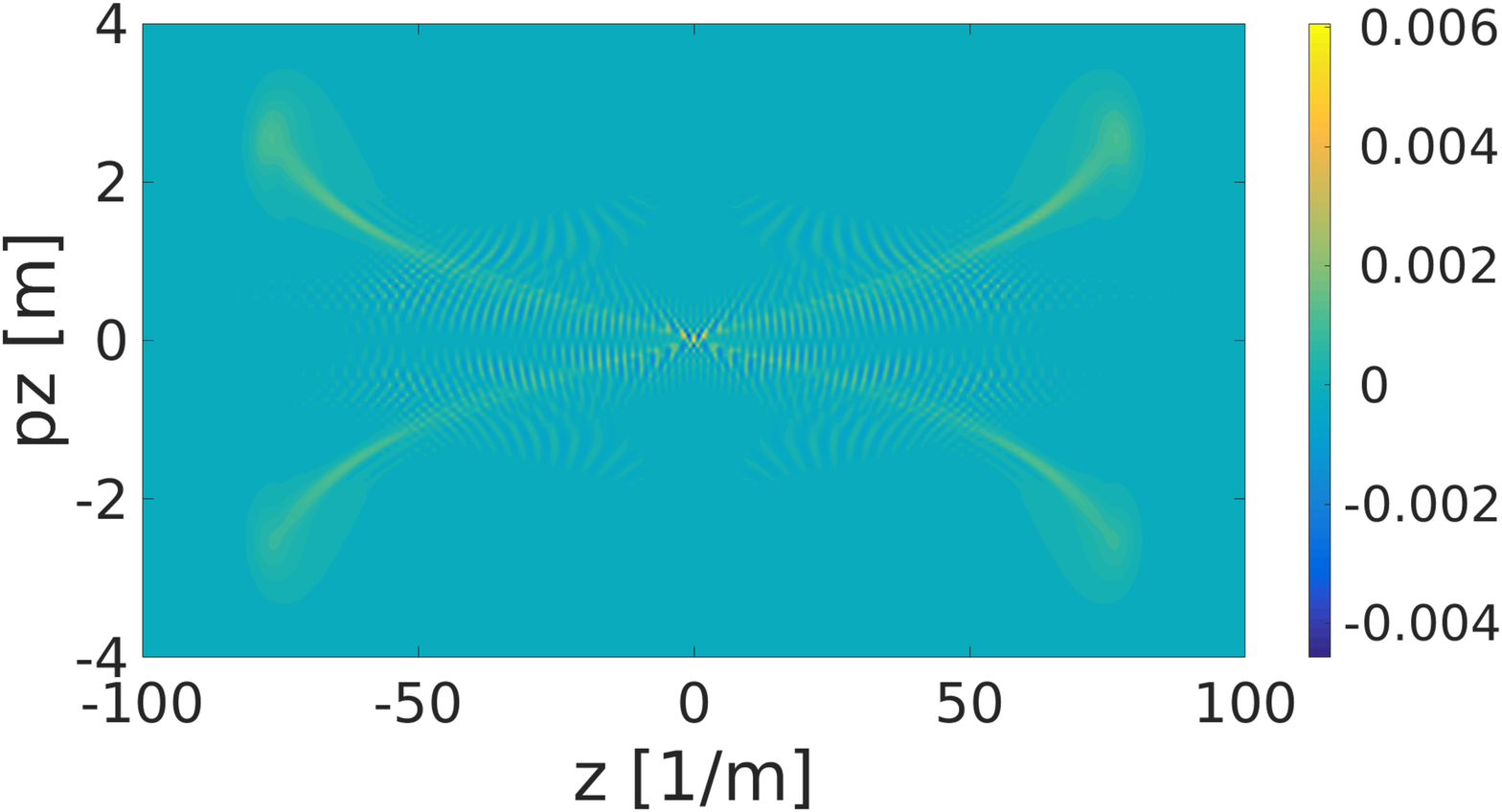} 
      \includegraphics[width=\figlen]{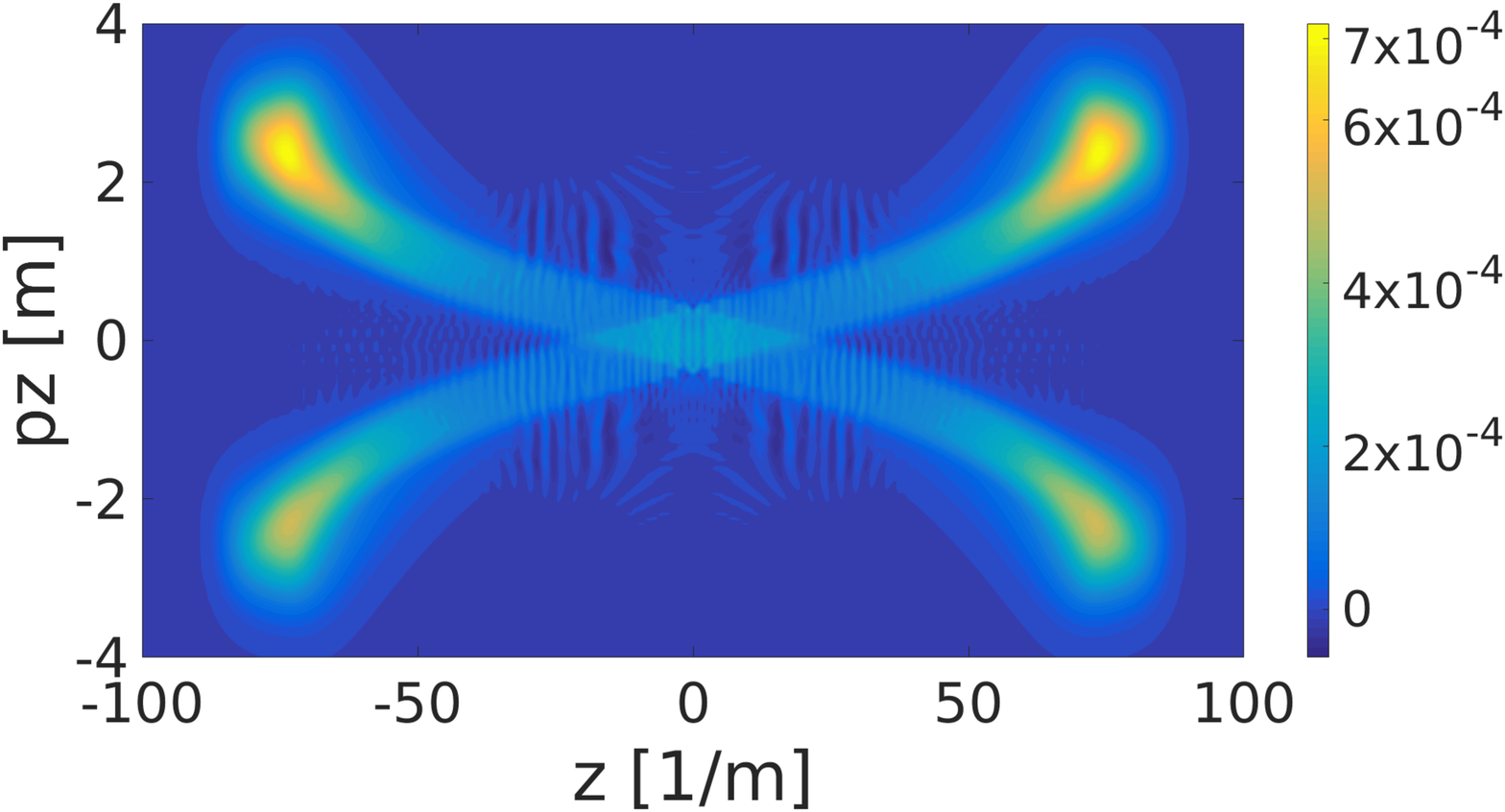}     
      \caption{Density plots of the particle distribution function $n \left( z, p_z \right)$ obtained from a DHW calculation at time $t_f=90 ~ m^{-1}$ without \txt{(left)}{(top)} and with 
        \txt{(right)}{(bottom)} averaging in phase space. Overall, the characteristic oscillatory pattern 
	of the Wigner distribution vanishes upon averaging, while signatures of particle creation remain intact. Parameters: $\varepsilon=0.5$,
	$\tau=20$ $m^{-1}$, $\lambda=10$ $m^{-1}$, $\omega=0.1$ $m$ and $\phi=0$ as well as $M=6$ and $\sigma=20$.}
      \label{fig:T20L10Blur}
      \end{figure}  

      \begin{figure}[t]
      \includegraphics[width=\figlen]{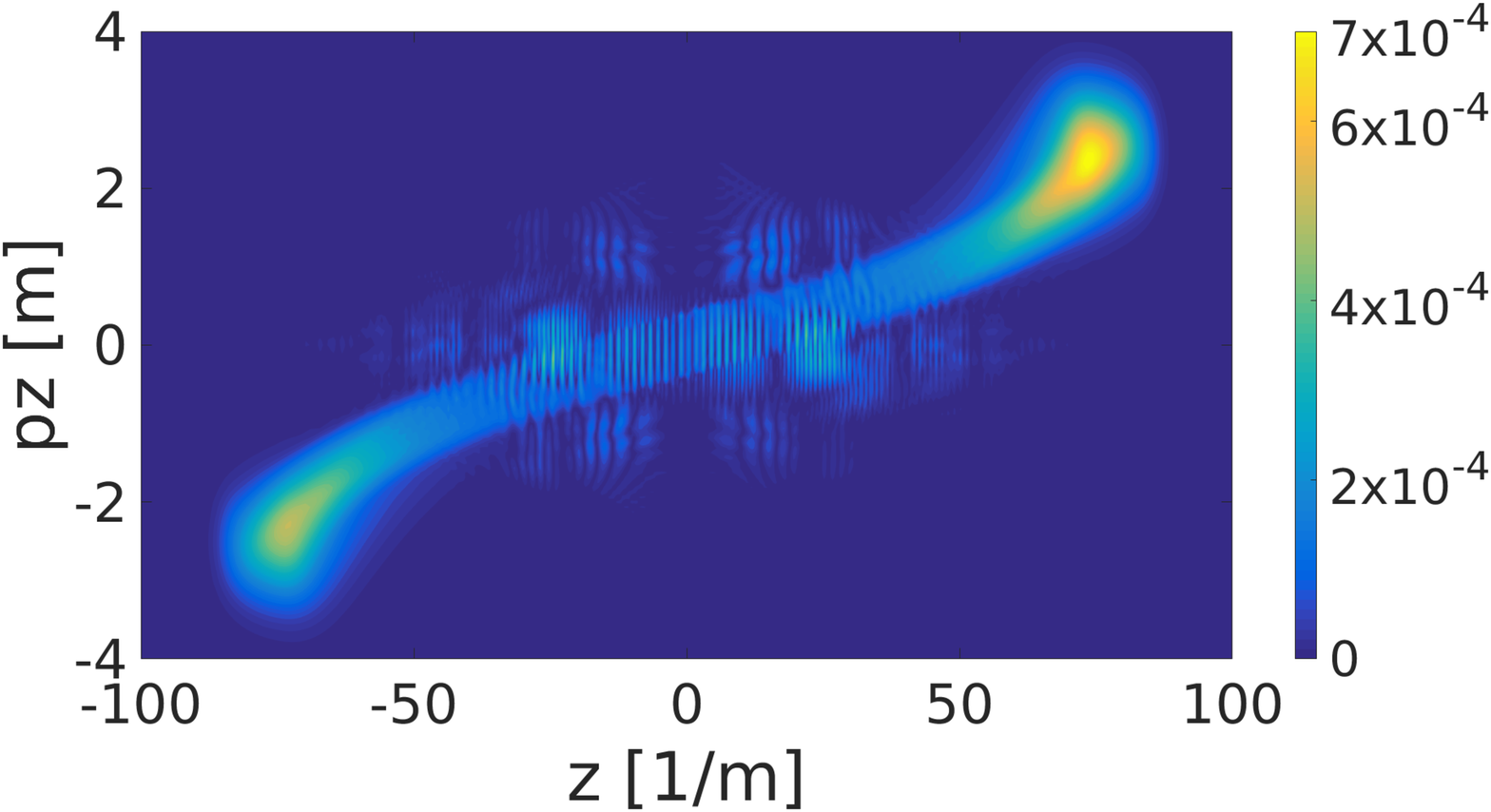} 
      \includegraphics[width=\figlen]{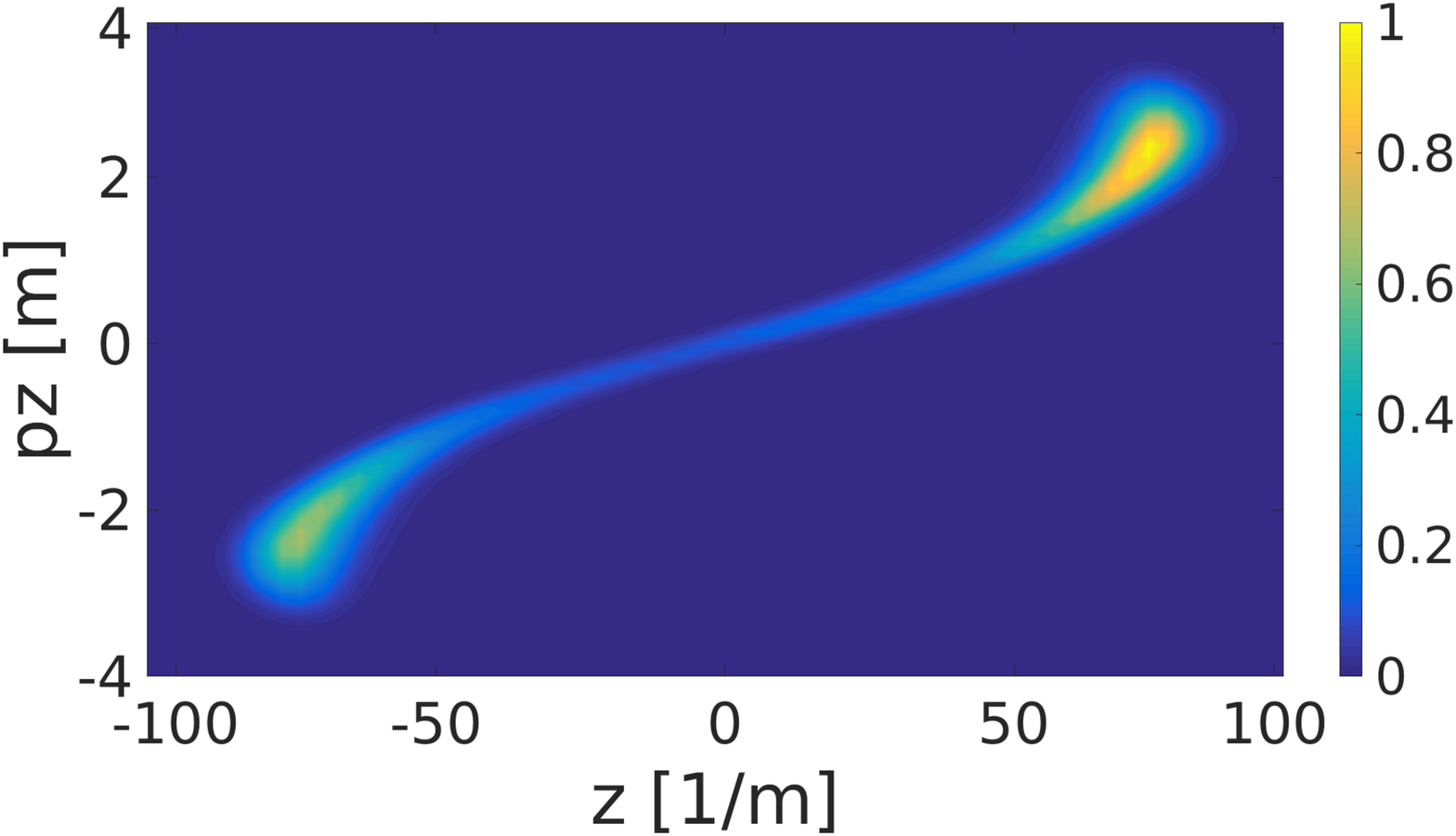}          
      \caption{Qualitative comparison of the absolute values for the coarse-grained positron distribution $f_{+} \left( z, p_z \right)$ \txt{(left)}{(top)}
        with a smooth single-particle positron density \txt{(right)}{(bottom)} where the spin $s=-1/2$. 
        The characteristic bulks as well as the asymmetric distribution are featured in both plots. 
	The structure at $p_z \approx 0$ in the first plot are remnants of the full Wigner function. Parameters: $\varepsilon=0.5$,
	$\tau=20$ $m^{-1}$, $\lambda=10$ $m^{-1}$, $\omega=0.1$ $m$, $t_f=90$ $m^{-1}$ and $\phi=0$ as well as $M=6$ and $\sigma=20$.}
      \label{fig:T20L10BlurE}
      \end{figure}  

In Fig.~\ref{fig:T20L10Blur} a comparison between the particle distribution function obtained from the DHW calculation before and after averaging is displayed. Discussing the direct output
from the DHW calculation we can discriminate two different structures: small smooth lines, that form an ``X'' and a rapidly oscillating pattern around $p_z \approx 0$.
This makes it obvious that an interpretation in terms of a probability density is not possible, because the distribution function yields
large negative values. Applying coarse graining procedures, we basically filter out these highly oscillatory parts. Although we inevitably lose information,
we gain a clearer picture of actually measurable quantities. Hence, apart from minor fluctuations (at the $10 \%$ level), we can in principle transform the quasi-probability
function into a probability density.

Fig.~\ref{fig:T20L10Blur} illustrates a further feature of the phase-space formalism. It seems as if four particle bulks are clearly separated from each other.
This picture, however, is misleading, because it displays particles and antiparticles as if they were sharing the same momentum variables, here $p_z$. This
is not correct in QFT, because antiparticles are generally defined with reverse momenta. Hence, in order to obtain a
meaningful result in terms of real particles one has to identify the proportion of the antiparticle distribution and reverse the signs
of the respective momentum coordinates. 

Given the phase-space particle distribution as well as the charge distribution we can discriminate particles from antiparticles. Let us write 
\begin{align}
 f_- \left( z, p_x, p_z \right) &= n \left( z, p_x, p_z \right)/2 - e c \left( z, p_x, p_z \right)/2, \\
 f_+ \left( z, p_x, p_z \right) &= n \left( z, p_x, p_z \right)/2 + e c \left( z, p_x, p_z \right)/2,
\end{align}
to account for the electron/positron distribution. Besides minor fluctuations, we observe in Fig.~\ref{fig:T20L10BlurE} the formation of
two bulks, which are, however, not equally dense. To understand this asymmetry, we discuss the results obtained from the trajectory model.

As we have derived the Wigner equations of motion \eqref{eqn1_1}-\eqref{eqn1_4} 
for one spin direction only, we have also fixed the spin $s$ in our model. With this point in mind, we are able to reproduce the particle distribution function in Fig.~\ref{fig:T20L10BlurE}.
The greatest strength of the trajectory model is, that it allows to examine the contributions of fields individually. Evaluating, for example, the semi-classical equations of motion \eqref{equ:sin1}-\eqref{equ:sin3} without
the spin-interaction 
term 
\begin{equation}
    \mathbf{F}_S = 
    \begin{pmatrix}
      0 \\
      \dfrac{s}{\gamma \left(t\right)} \left( \partial_z B \left(t,z(t) \right) - \frac{p_z \left(t\right)}{\gamma \left(t\right)+1} \partial_z E \left(t,z(t) \right) \right) 
    \end{pmatrix}
\end{equation}
yields a perfectly symmetric particle distribution. In the next step, we performed two additional calculations, where we switched off the term $\partial_z B \left(t,z(t) \right)$ in scenario (i)  
and the term $\partial_z E \left(t,z(t) \right)$ in scenario (ii). In the latter, we could nicely reproduce the results given in Fig.~\ref{fig:T20L10BlurE}, thus we conclude that  
the magnetic field gradient is mainly responsible for breaking the $p_z$-symmetry.        
      
\subsection{Particle distribution in momentum space}
\label{ch:MomSpace} 
            
\begin{figure}[t]
      \begin{center}
	\includegraphics[width=\figlen]{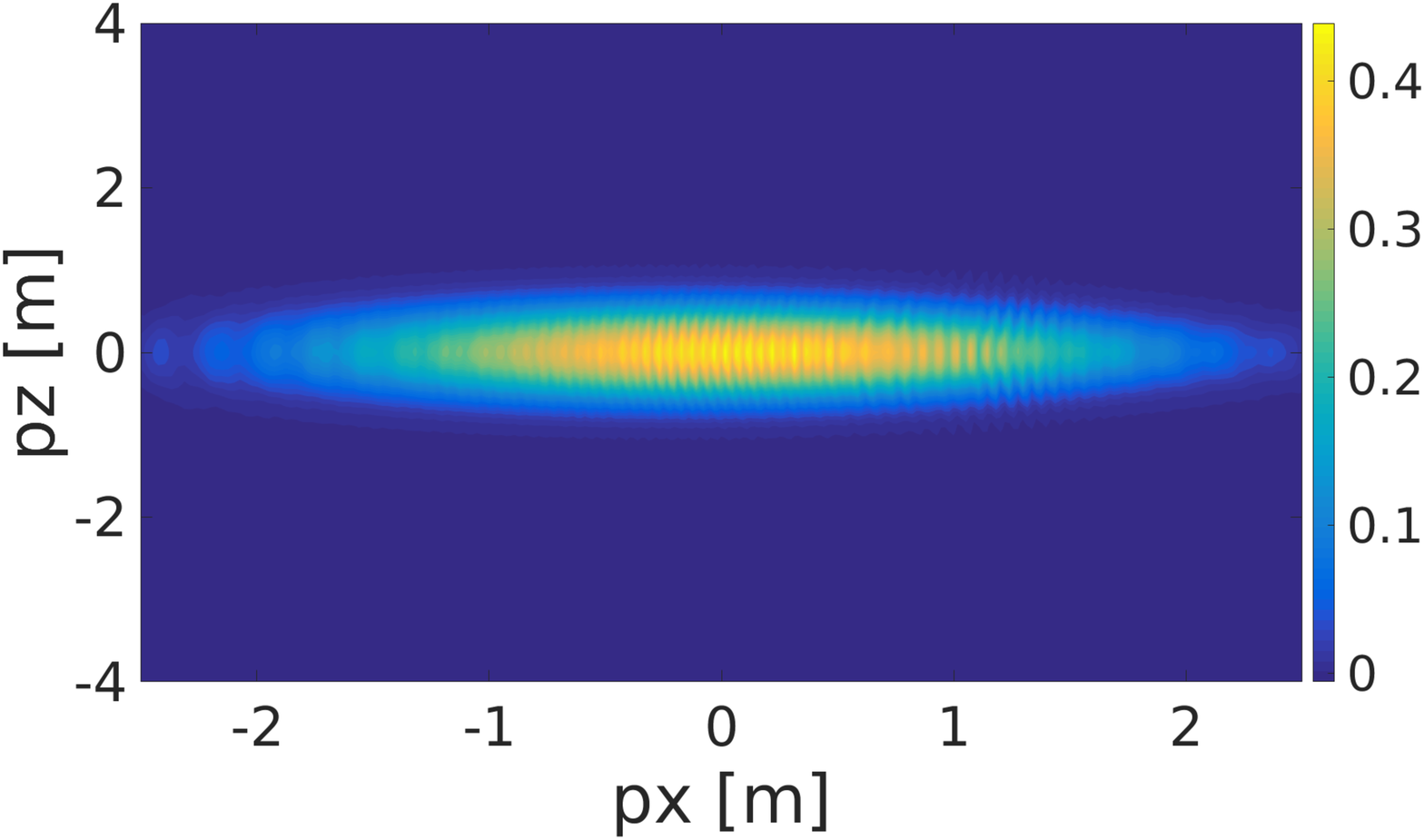}       
	\includegraphics[width=\figlen]{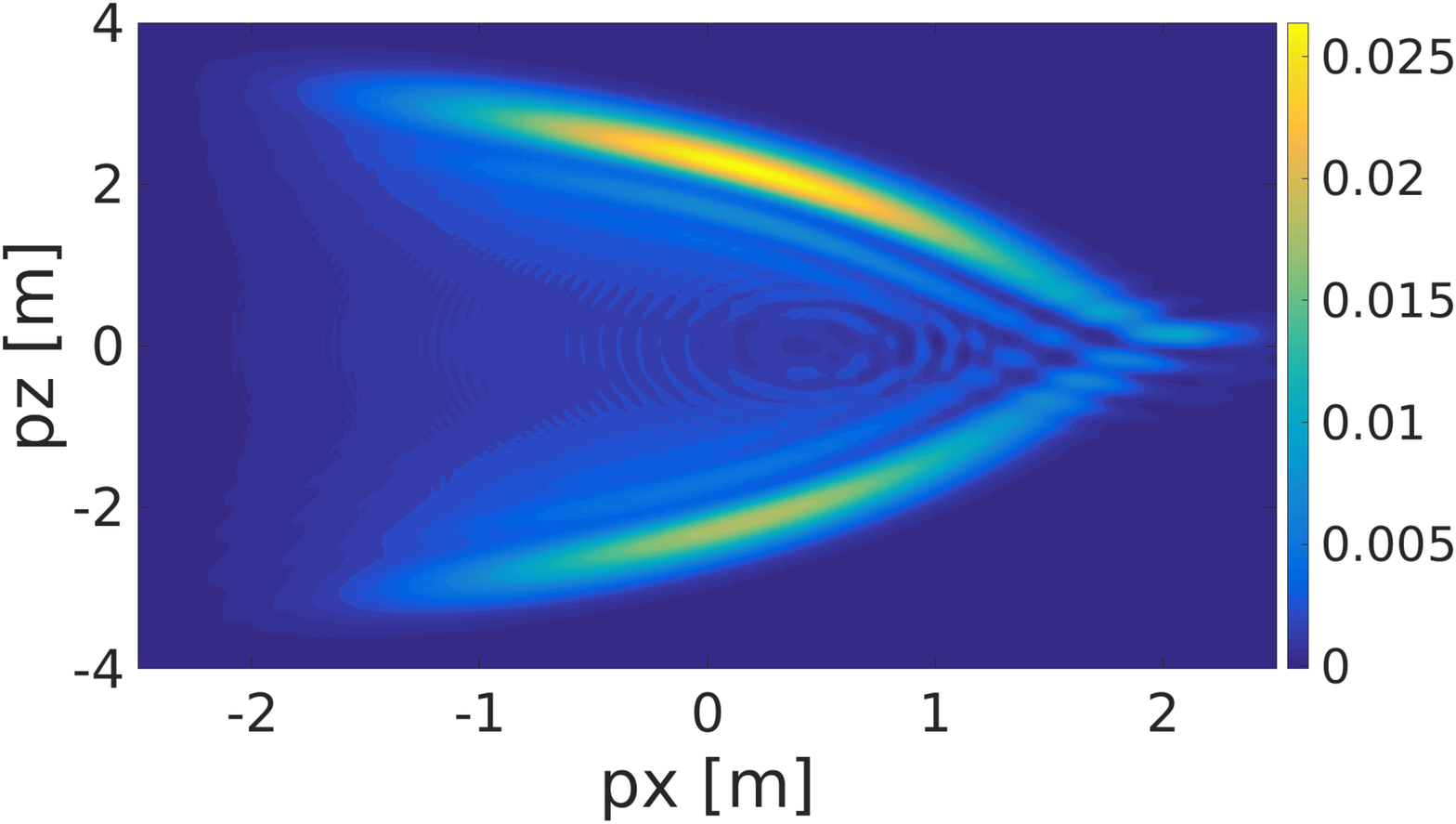}		
      \end{center}
      \caption{Density plots of the particle distribution function $n \left( p_x, p_z \right)$ (DHW calculation) 
        in momentum space for $\varepsilon=0.5$, $\tau=20$ $m^{-1}$, $\omega=0.1$ $m$, $\phi=0$ 
	and $\lambda=100$ $m^{-1}$ \txt{(left)}{(top)} or $\lambda=10$ $m^{-1}$ \txt{(right)}{(bottom)}. 
	Particles created at early times are strongly accelerated towards negative momentum $p_x$ by the main peak of the electric field and then deflected in
	$z$-direction by the magnetic field. Particles created at late times only acquire substantial momentum (towards positive $p_x$), because of the minor peak in the electric field.
	Due to the presence of (i) a favored spin direction and (ii) a strong magnetic field gradient, the asymmetry in $p_z$ can be attributed to direct spin-field interactions.}
      \label{fig:T20L10}
      \end{figure}

Signatures of a spin-field interaction show up also in the distribution functions in momentum space. Solving the DHW equation \eqref{eqn1_1}-\eqref{eqn1_4} for large spatial
extent ($\lambda=100$ $m^{-1}$) and thus weak magnetic field leads to a distribution that is strongly confined in $p_z$-direction, see Fig.~\ref{fig:T20L10} (top).       
     
With the aid of the semi-classical model, the particle distribution in Fig.~\ref{fig:T20L10} can be very well understood assuming that
particles are created with vanishing initial longitudinal momentum, but finite transversal momentum. 
\ck{
Employing a strong, quasi-homogeneous short-pulsed field ($\varepsilon=0.5$, $\tau=20$ $m^{-1}$ and $\lambda=100$ $m^{-1}$) we can assume that neither the particle position $z(t)$ nor
the magnetic field $B \left(t,z(t) \right)$ play any significant role. Hence, we can assume that for this field configuration the force terms $\mathbf{F}_B$ and $\mathbf{F}_S$ in
the equations of motion Eqs. \eqref{equ:sin2} - \eqref{equ:sin3} vanish. As a result the equations of motion take the simple form
\begin{alignat}{3}
 & \frac{\partial p_x}{\partial t} &&= E \left(t \right),&& \qquad p_x(t_0) = 0, \\
 & \frac{\partial p_z}{\partial t} &&= 0,&& \qquad p_z(t_0) = p_{z,0}.
\end{alignat}
In such a case it is possible to solve the differential equation analytically yielding
\begin{align}
 p_x(t) &= A(t_0) - A(t), \\
 p_z(t) &= p_{z,0} = {\rm const}.
\end{align}
Due to the fact, that the vector potential vanishes at asymptotic times the final particle momentum therefore solely depends on the vector potential at the particles time of creation.
To be more specific, the peak in Fig.~\ref{fig:T20L10} (top) at $p_x = 0$ is due to the dominant peak in the electric field at $t=0$. As $E(t)$ decreases at later times less
particles are produced. However, as $A(t)$ increases at the same time, these particles are effectively accelerated much stronger. Due to the symmetry of the field with respect to $t$ the same is true
for $t < 0$.}


One advantage of the DHW formalism is, that it allows to study pair production in strongly varying, spatially inhomogeneous fields. An example is given in Fig.~\ref{fig:T20L10}, 
where $\lambda=10$ $m^{-1}$ and $\omega=0.1$ $m$. This configuration yields, as a side effect, also a strong magnetic field. Again, particles are created around the main peak of the electric field and
subsequently accelerated due to the strong force on the charged particles. In turn, depending on the time of creation, the particles acquire a different momentum $p_x$.
The earlier a particle pair is created the more it is accelerated by the electric field (here in negative $x$-direction). However, by contrast with the previous consideration, here
the magnetic field is strong enough to transfer a substantial amount of momentum from $p_x$ to $p_z$, \ck{c.f. the force term
\begin{equation}
    \mathbf{F}_B = 
    \begin{pmatrix}
      -  \frac{p_z \left(t\right)}{\gamma \left(t\right)} B \left(t,z(t) \right) \\
      + \frac{p_x \left(t\right)}{\gamma \left(t\right)} B \left(t,z(t) \right)
    \end{pmatrix}
\end{equation}
in Eqs. \eqref{equ:sin2}-\eqref{equ:sin3}.}
Such a conversion of momentum is not possible for particles produced at later times
($p_x$ stays close to zero at first). Due to the fact, that these particles are nearly unaffected by the magnetic field they cannot be pushed away from the strong field region.
In turn, they are basically only accelerated by the second minor peak in the electric field (in positive $x$-direction).
Hence, the ``$>$'' shaped structure of the particle distribution.  

Additionally, the second peak in time of the magnetic field shows a strong spatial gradient.
Hence, the magnetic field directly interacts with the particle spin, similarly to the Stern-Gerlach experiment. 
\mbox{(Anti-)}fermions have non-zero spin, thus depending on their spin alignment they feel an additional
force. As we performed calculations for only one spin-direction, this force pushes particles only in one $p_z$-direction. 
Eventually, analyzing a collection of particles, this leads to a net force breaking $p_z$-symmetry.

      \begin{figure}[t]
      \begin{center}
	\includegraphics[width=\figlen]{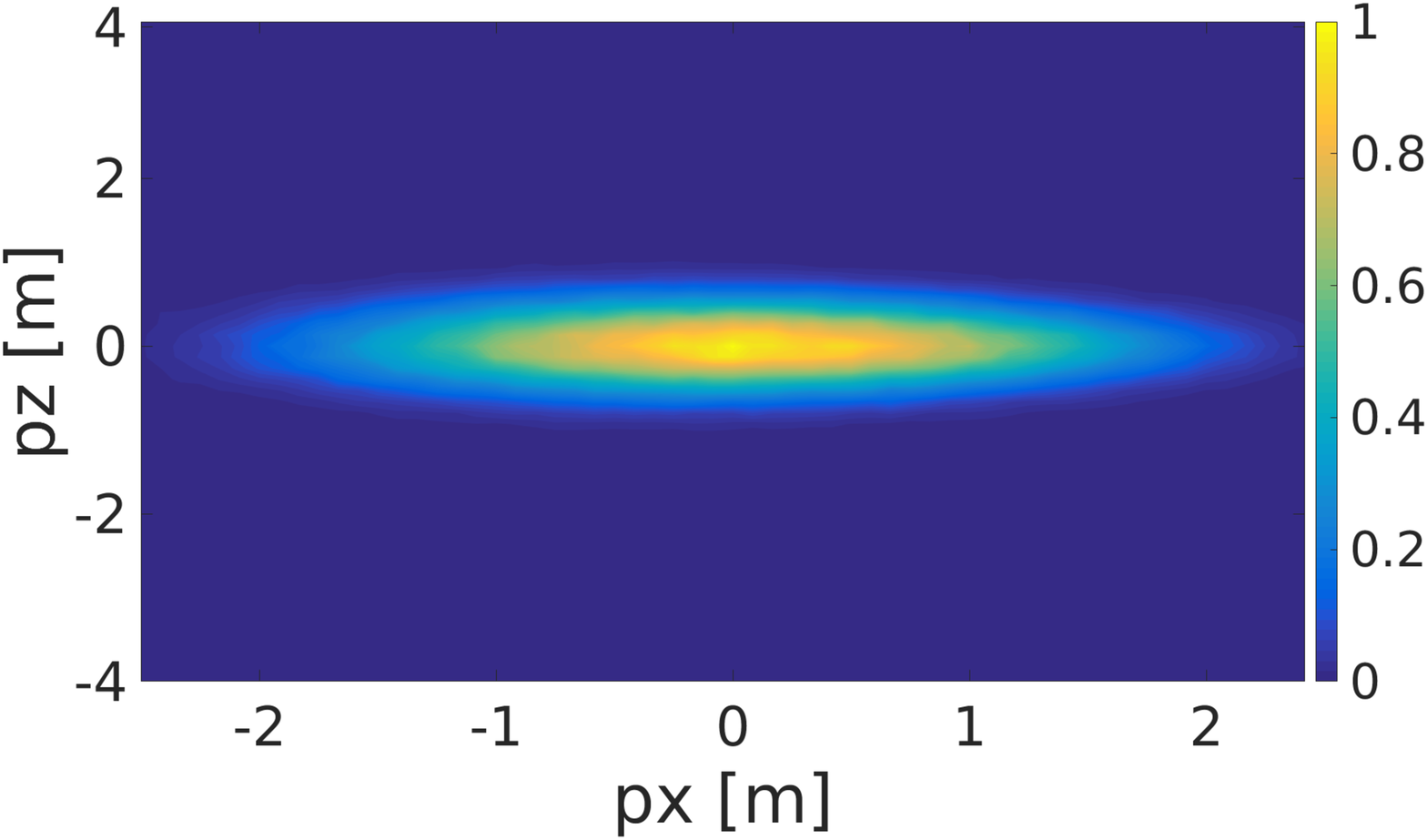} 
	\includegraphics[width=\figlen]{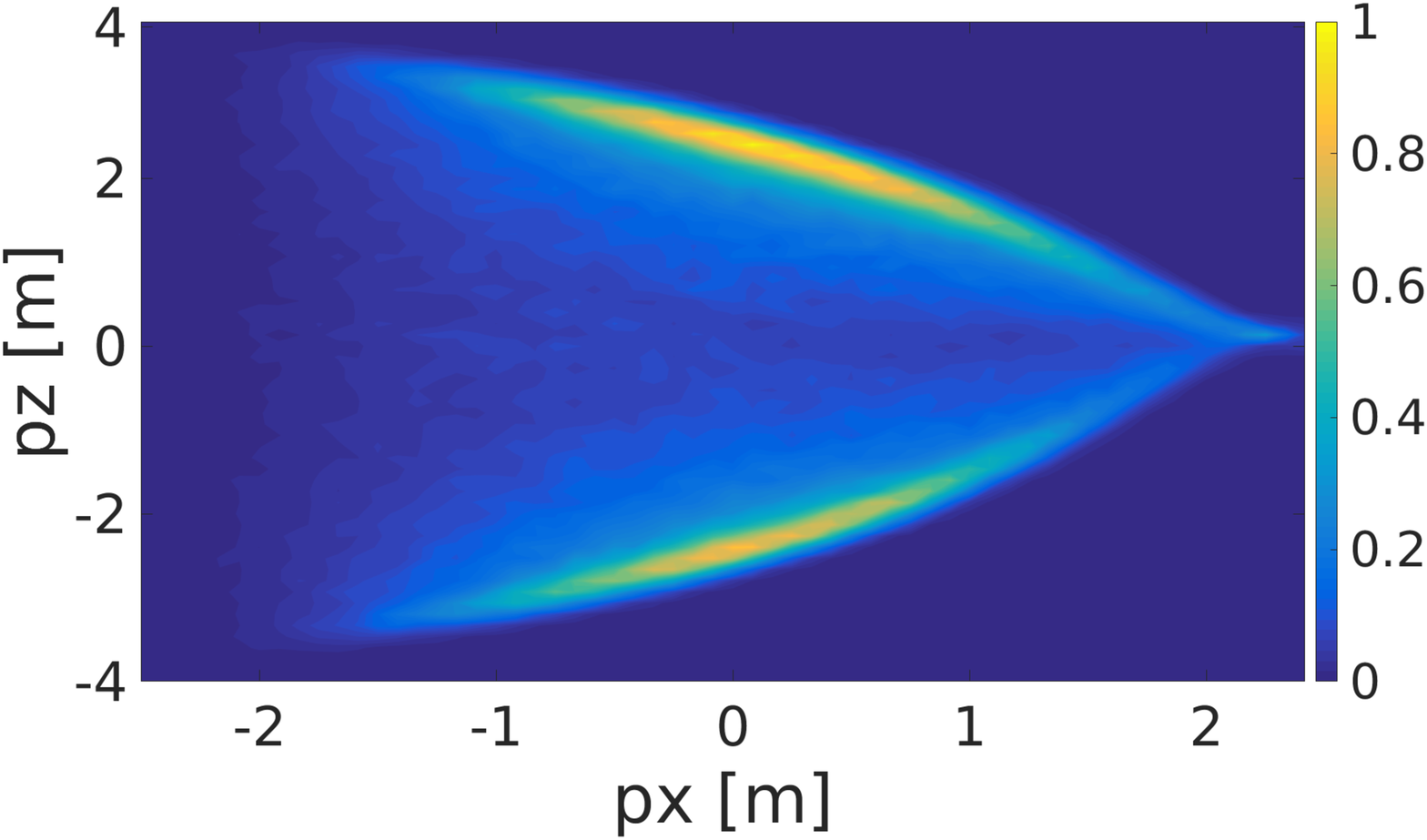} 
      \end{center}
      \caption{Smooth density histogram of the single-particle distribution as function of the momenta $\left( p_x, p_z \right)$ at final time $t_f=90$ $m^{-1}$. 
	A total of approximately $500,000$ trajectories have been evaluated, interpolated and normalized. The stronger the magnetic field the more momentum can be transferred from $p_x$
	to $p_z$. In turn, the particle bulk becomes widely distributed in $p_z$-direction. Background field parameters: $\varepsilon=0.5$, $\tau=20$ $m^{-1}$, $\omega=0.1$ $m$, $\phi=0$, 
	$\lambda=100 ~ m^{-1}$ \txt{(left)}{(top)} or $\lambda=10$ $m^{-1}$ \txt{(right)}{(bottom)}.}      
      \label{fig:T20L10Traj}
      \end{figure}   

We have illustrated the smooth particle densities on the basis of the trajectory model in Fig.~\ref{fig:T20L10Traj}. Studying trajectories essentially supports
our interpretation of \mbox{Fig.~\ref{fig:T20L10}}. However, in contrast to the semi-classical analysis a computation based on the DHW formalism describes quantum interferences.
This explains the differences in Figure \ref{fig:T20L10} and Figure \ref{fig:T20L10Traj}. The DHW results show a strong interference pattern, visible as additional
side maxima. This feature is clearly missing in the semi-classical calculations, where one obtains only the peaks without any interference pattern.       

\subsection{The envelope phase}
\label{ch:Phase}

\begin{figure}[t]
      \begin{center}
	\includegraphics[width=\figlen]{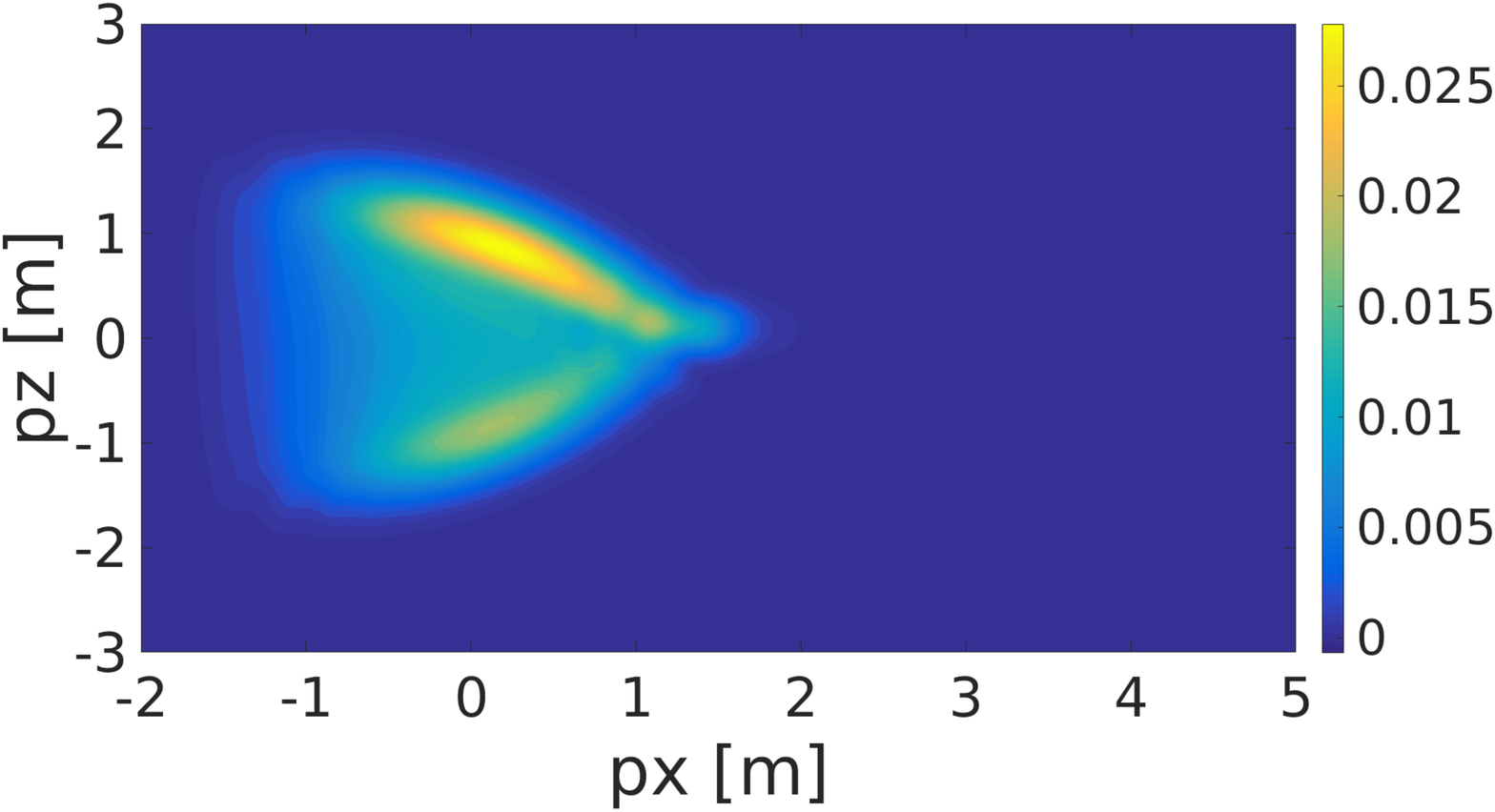} 
	\includegraphics[width=\figlen]{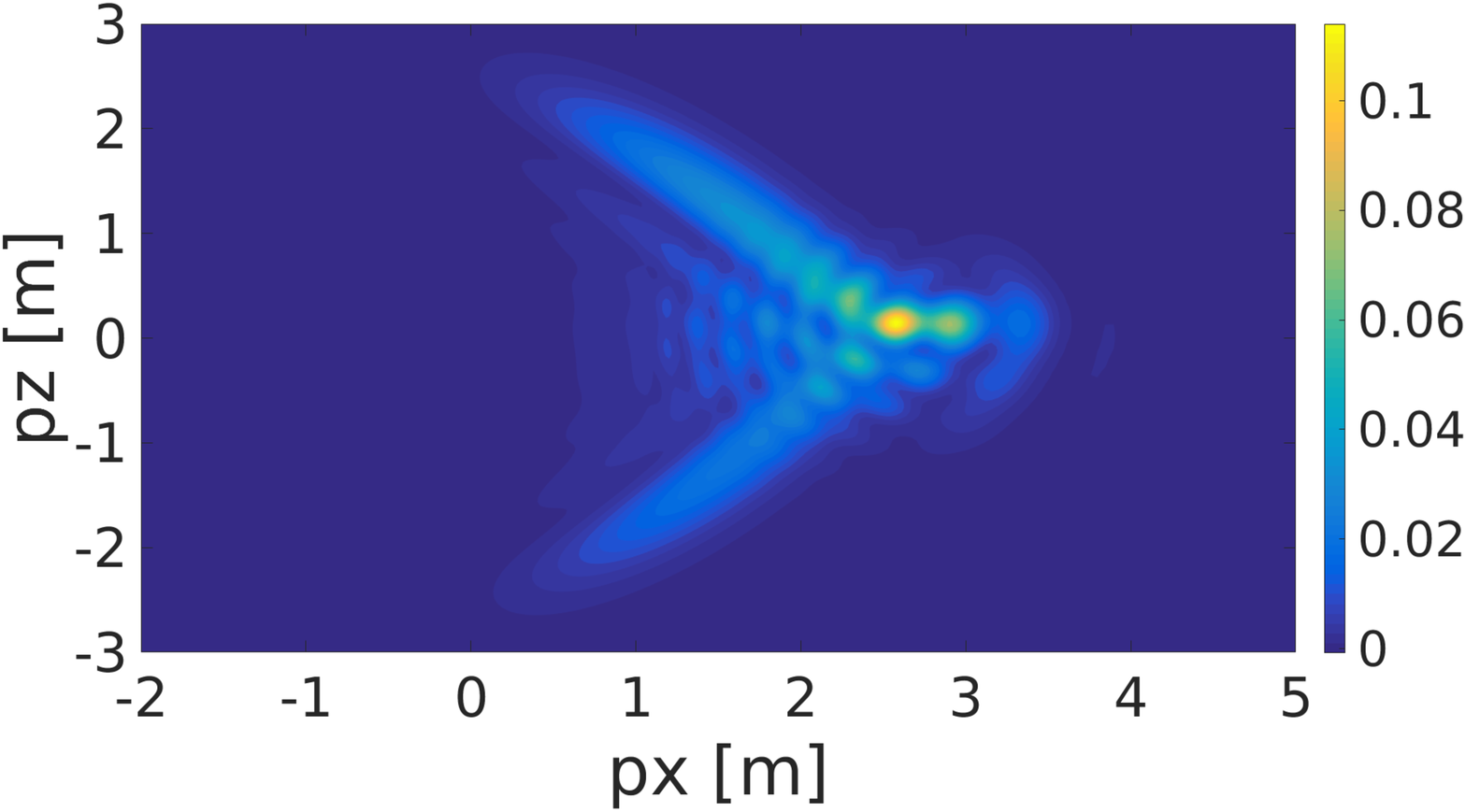} 
	\includegraphics[width=\figlen]{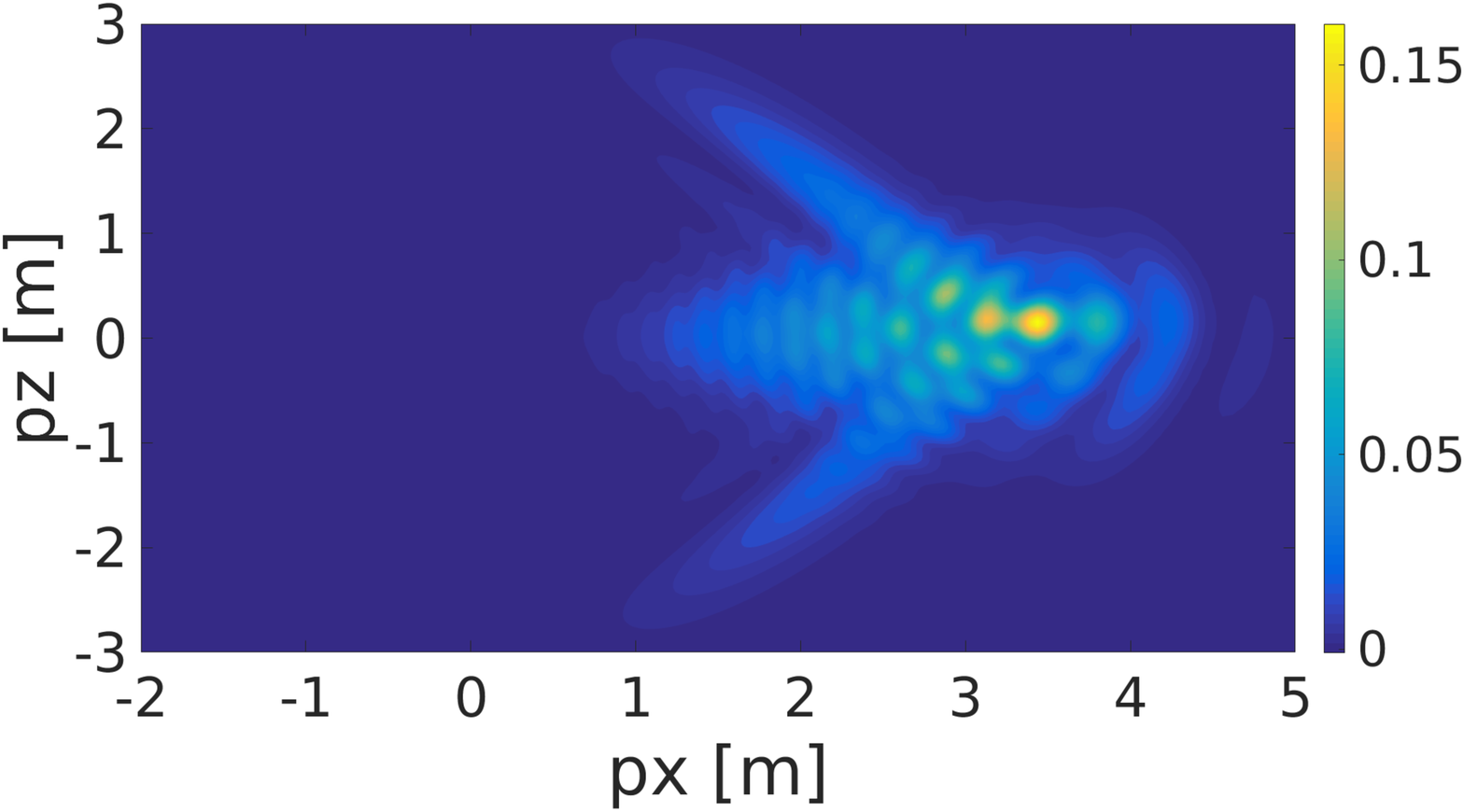} 
	\includegraphics[width=\figlen]{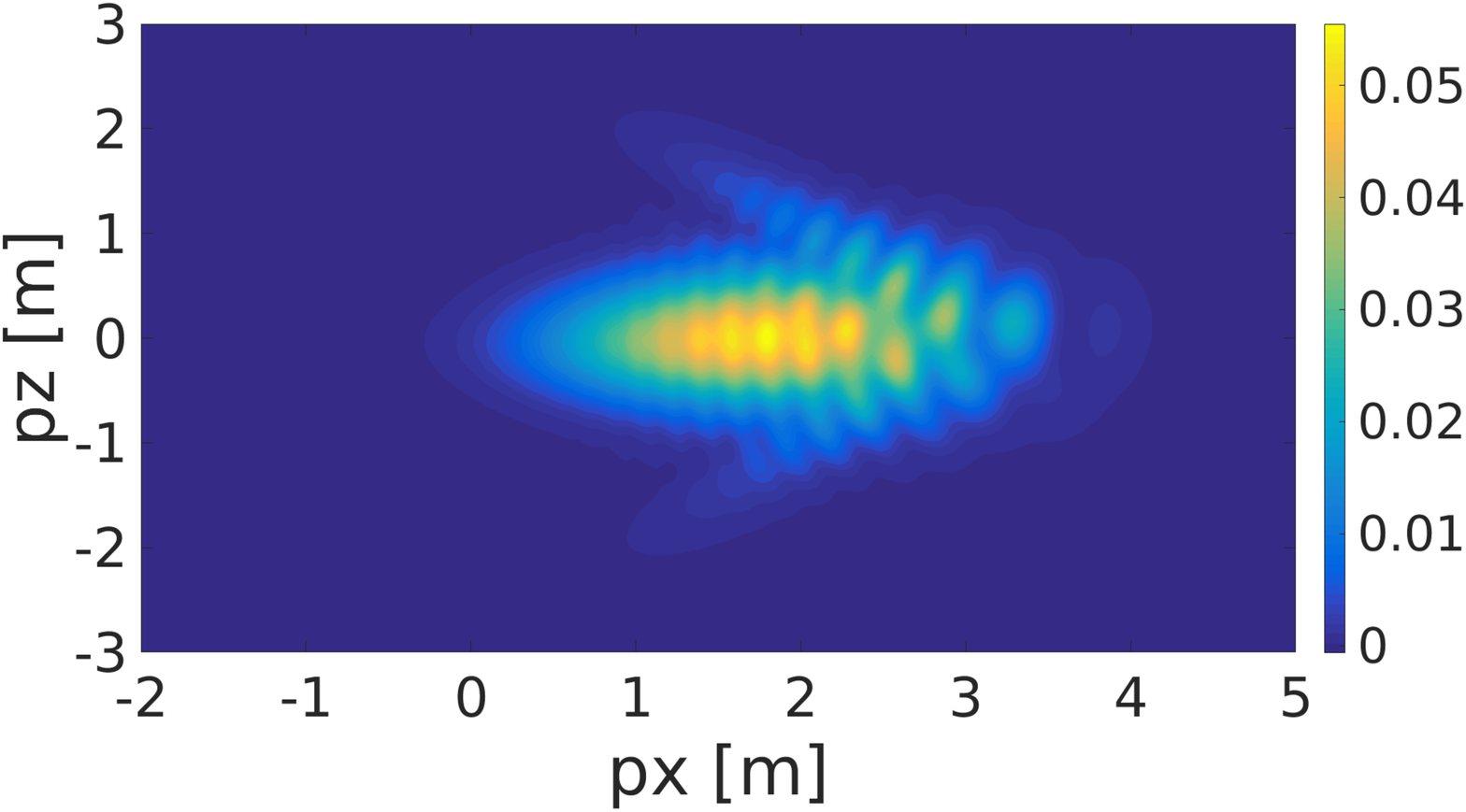} 		
      \end{center}
      \vspace{-0.1cm} 
      \caption{Series of density plots of the particle distribution function $n \left( p_x, p_z \right)$ obtained from a DHW calculation. 
        The envelope phase is varied \txt{}{(top to bottom)};
	\txt{$\phi=0$ (top left),}{$\phi=0,$} \txt{$\phi=\pi/4$ (top right),}{$\phi=\pi/4,$} \txt{$\phi=\pi/2$ (bottom left),}{$\phi=\pi/2,$} \txt{$\phi=3\pi/4$ (bottom right)}{$\phi=3\pi/4,$} for fixed 
	$\varepsilon=0.5$, $\tau=10$ $m^{-1}$, $\omega=0.1$ $m$ and $\lambda=10$ $m^{-1}$.
	The interference pattern is a characteristic feature of pair production in electric fields with multiple prominent peaks.}
      \label{fig:T10L10}
      \end{figure}

In the following, we apply our findings to field configurations exhibiting more than one prominent peak in the electric field. 
Solving the DHW equation \eqref{eqn1_1}-\eqref{eqn1_4} for various $\phi$, we are able to display a series of density plots in Fig.~\ref{fig:T10L10} 
demonstrating the influence of the envelope phase in a few-cycle pulse. The first picture illustrates the particle
density for a field configuration with one dominant peak in the electric field (general parameters: $\phi=0$, $\lambda=10$ $m^{-1}$ and $\tau=10$ $m^{-1}$).
Due to the shorter interaction time compared to the previous discussion, see Sec. \ref{ch:MomSpace}, quantum interferences play only a minor role. The overall particle distribution, however,
is qualitatively the same.  

The picture changes drastically when considering a non-vanishing envelope phase $\phi$. 
As already investigated in Refs. \cite{Akkermans:2011yn,Hebenstreit:2009km}, an electric field exhibiting multiple peaks in time can act as a double-slit experiment. 
The same seems to be true for spatially inhomogeneous fields. Particles created at different peaks in time can still occupy the same phase-space volume, 
thus leading to quantum interferences. 

      \begin{figure}[ht]
      \begin{center}     
	\includegraphics[width=\figlen]{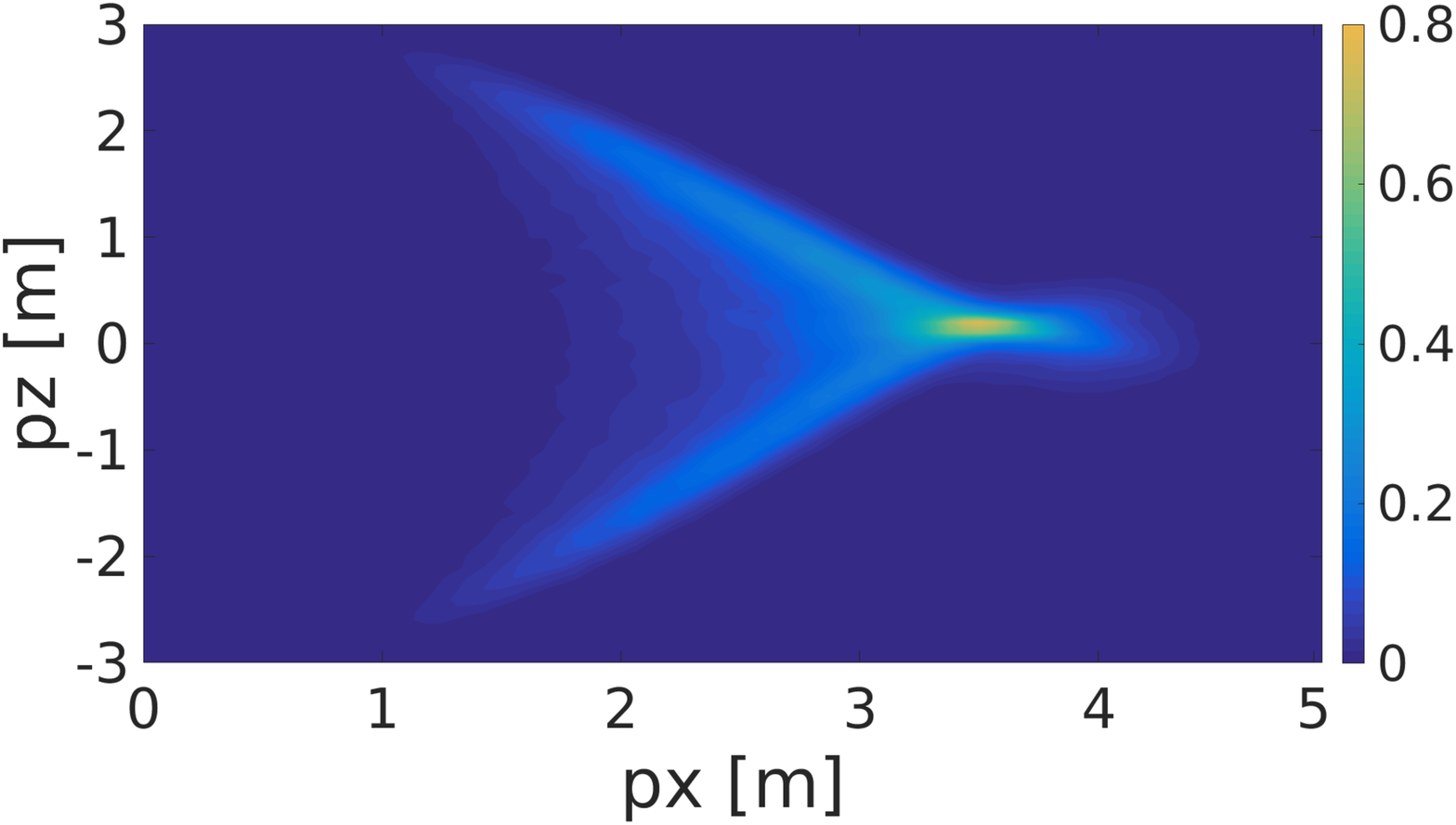} 
	\includegraphics[width=\figlen]{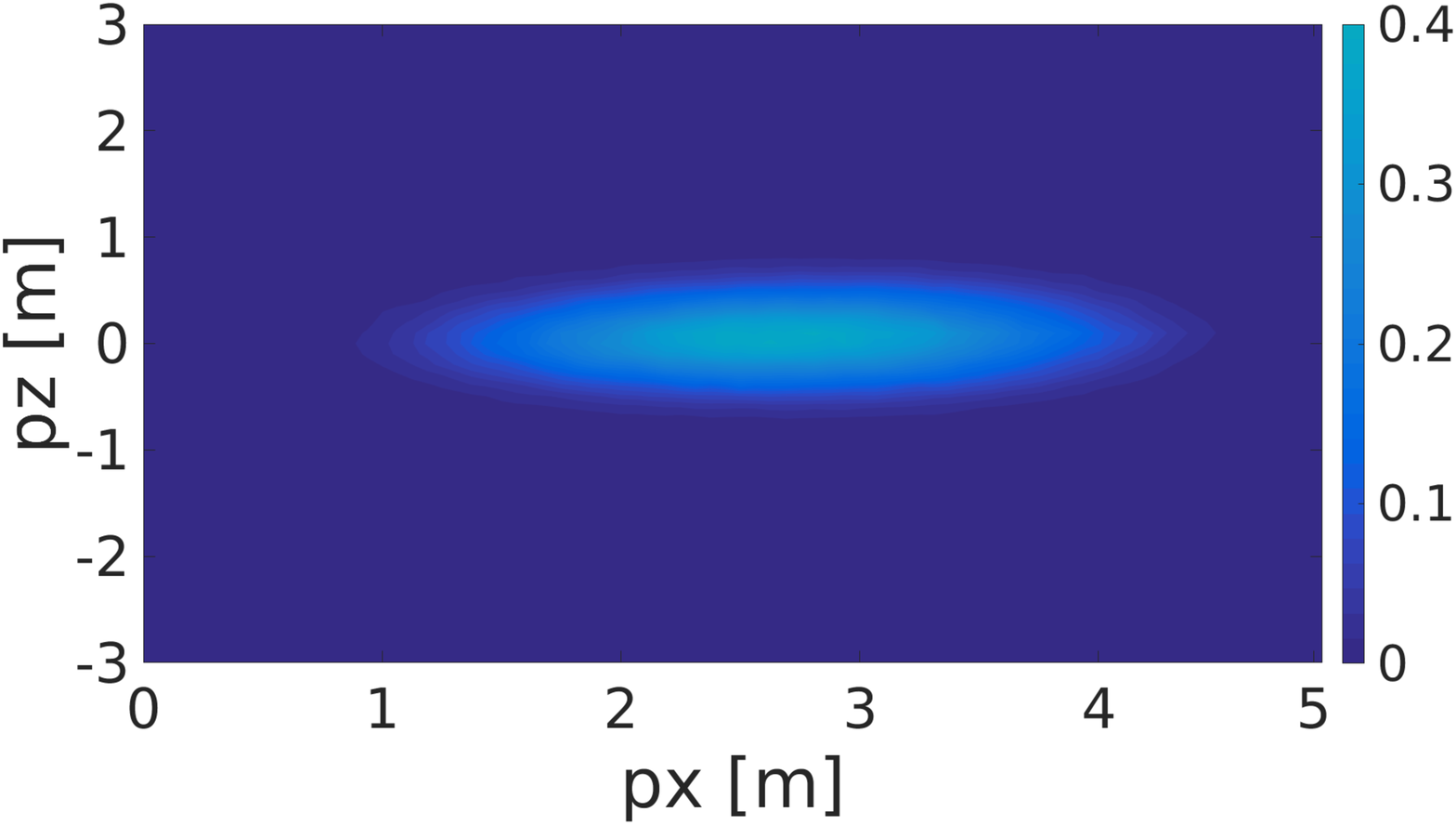} 
	\includegraphics[width=\figlen]{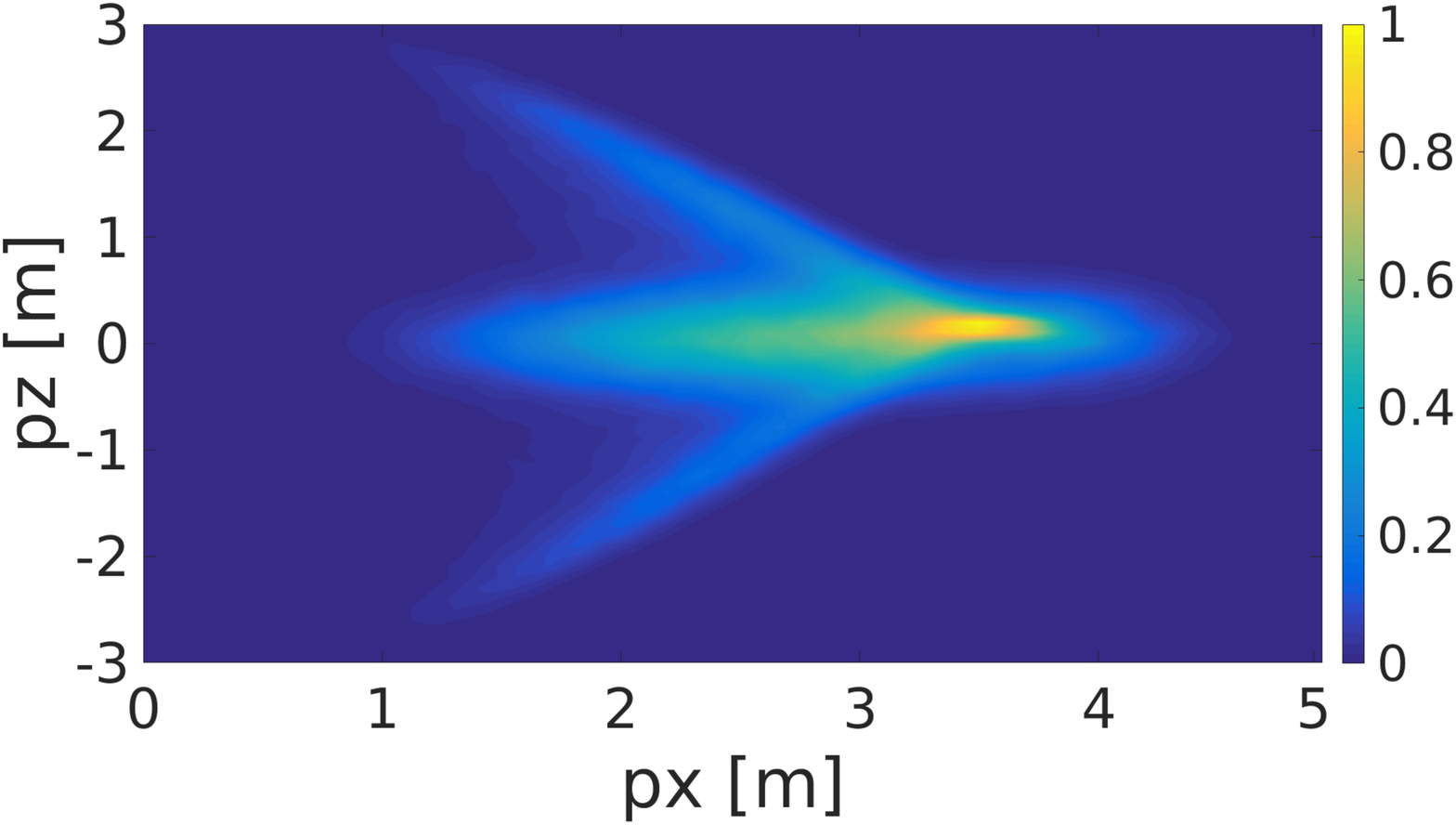} 	
      \end{center}
      \caption{Smooth density histogram of the single-particle distribution within a semi-classical trajectory analysis 
        broken down into contributions from the first \txt{(left)}{(top)}, the second (middle) and both \txt{(right)}{(bottom)} peaks in the electric field for $\phi=\pi/2$.
	Particles produced at the first peak are greatly accelerated and dispersed by the magnetic field. Nevertheless, both peaks produce the same total number of particles. Background field:
	$\varepsilon=0.5$, $\tau=10$ $m^{-1}$, $\omega=0.1$ $m$ and $\lambda=10$ $m^{-1}$.}
      \label{fig:T10L10Inter}
      \end{figure}    

Choosing, for example, a phase of $\phi=\pi/4$ results in an electric field, that features one prominent and one smaller peak in time.
According to the semi-classical model, the greater peak at $t < 0$ produces a larger amount of particles. These
particles, however, are exposed to a strong magnetic field leading to the ``wings'' in the particle density. Nevertheless, a substantial
fraction of particles is nearly unaffected by the magnetic field. These particles, together with the particles created at the second peak,
eventually culminate in a distribution function superimposed by wave peaks and troughs.
The most prominent example, however, is given in case of $\phi=\pi/2$, where both extrema in the electric field show the same absolute value.
Although the magnetic field still deflects many particles the interference pattern dominates the distribution.   

For the sake of completeness, we have also
displayed the result of a configuration with $\phi=3\pi/4$ in Fig.~\ref{fig:T10L10}. Basically, the situation is opposite to the case with \mbox{$\phi=\pi/4$}. Now, most particles are created after
the magnetic field vanished, thus they are only affected by the electric field. As a result the distribution is more confined and the quantum interferences play only
a minor role. Note, that we have neither normalized the distribution function nor the electric peak field strength. It is therefore not surprising, that the peak numbers change drastically.      

The trajectory analysis approach makes the perfect tool in order to analyze the particle distributions in Fig.~\ref{fig:T10L10}. 
To support our interpretation we have therefore exemplarily evaluated a total of
approximately $500,000$ trajectories for a configuration with $\phi=\pi/2$ within the semi-classical approach, see Fig.~\ref{fig:T10L10Inter} for an illustration.
The electric background field is antisymmetric in time, see Fig.~\ref{fig:EB2}, thus particle production mainly takes place around the two different peak positions. If particle pairs
are created at the first peak in time, they are exposed to very strong electric and magnetic fields, where especially the latter is accelerating particles into the $z$-direction.
By contrast, the magnetic field has nearly vanished at the time the second peak in the electric field starts to produce particles. As a result, nearly no conversion in momentum
takes place.
Although the particles are distributed differently, both bulks contribute evenly towards the total production rate. This is not surprising, because the peaks in the electric field
differ only by a sign.

     \begin{figure}[t]
      \begin{center}
	\includegraphics[width=\figp]{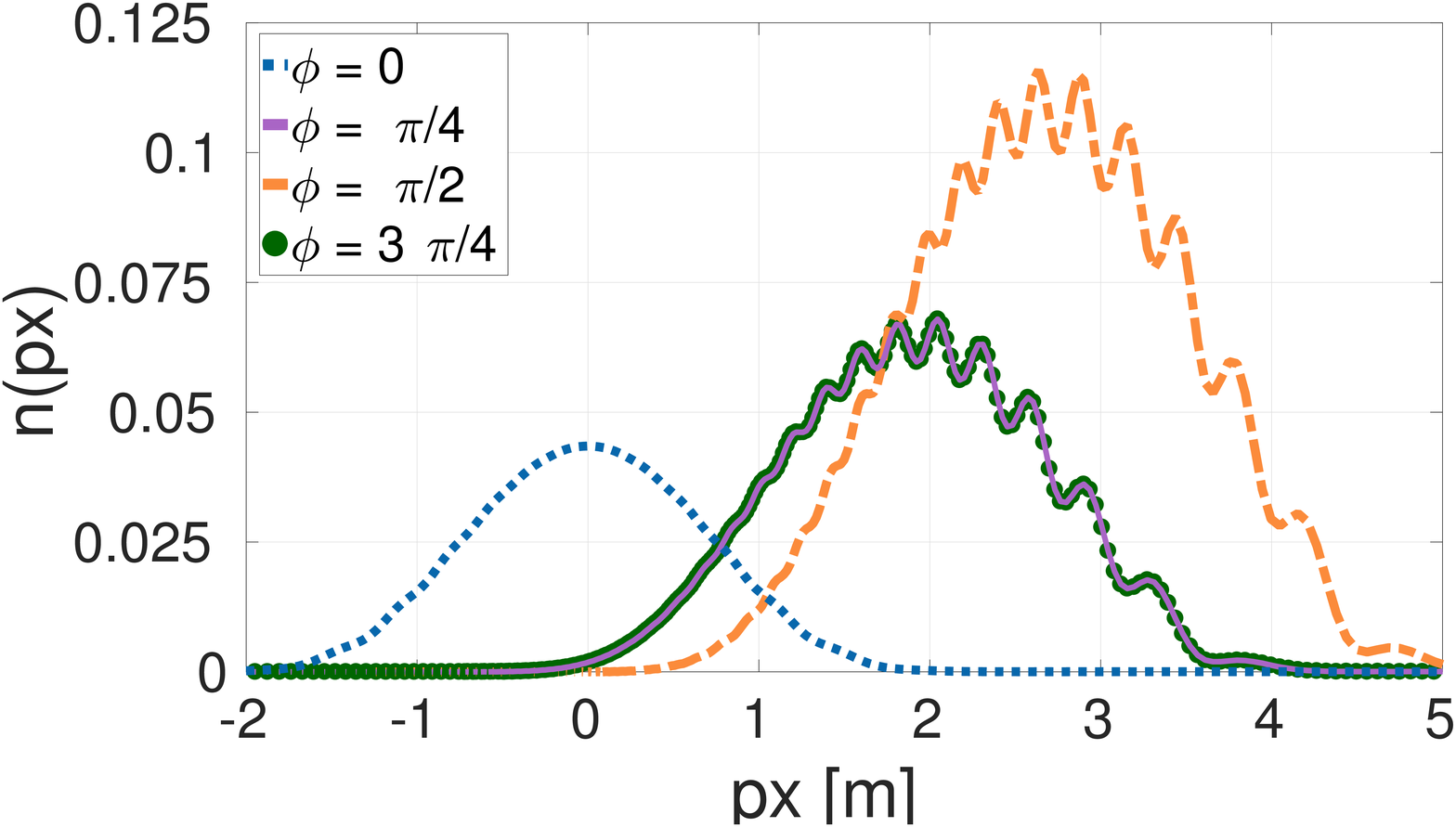} 
	\includegraphics[width=\figp]{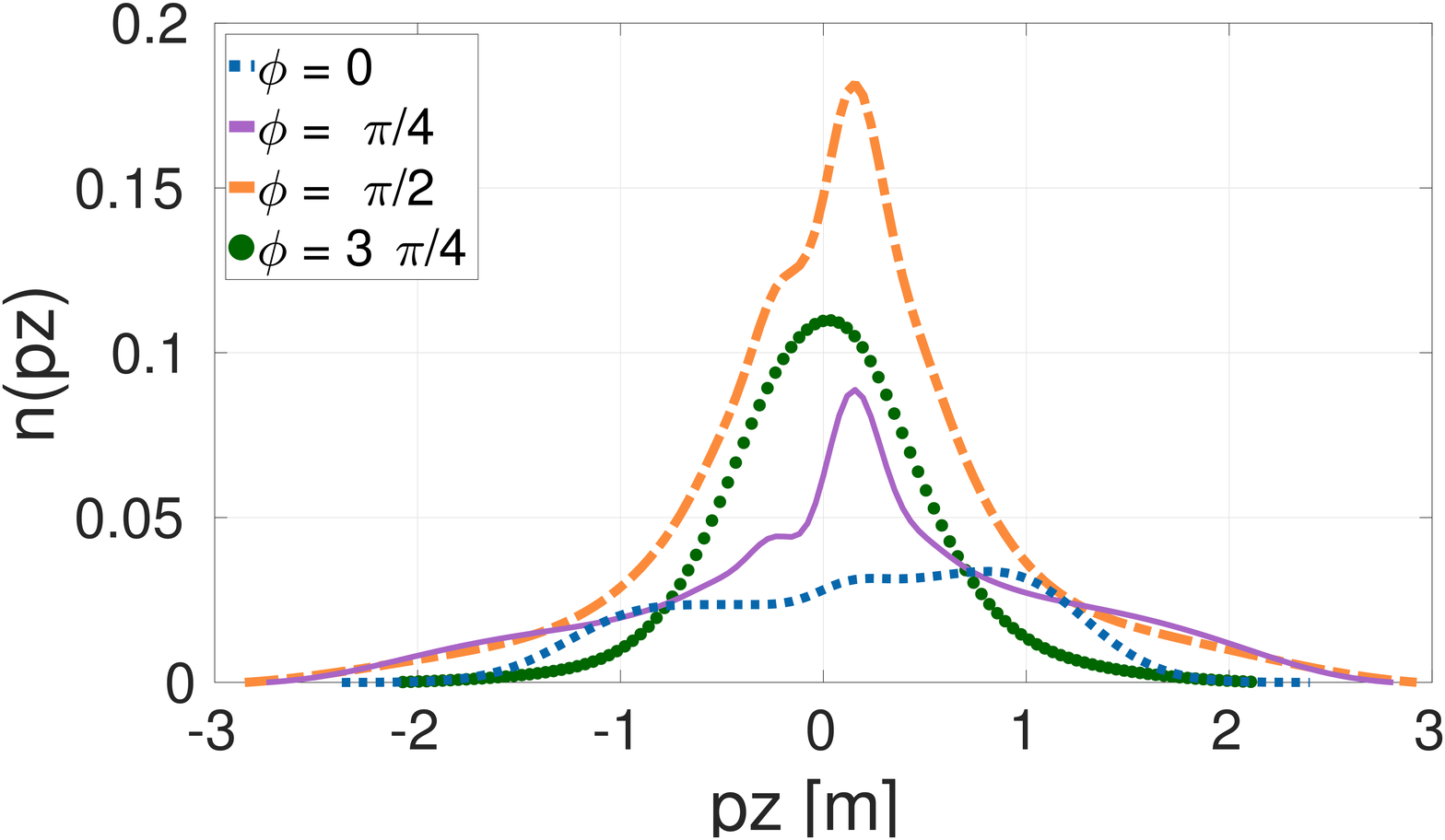} 
      \end{center}
      \caption{Particle spectra for Schwinger pair production as function of the momenta $p_x$ and $p_z$. The interference pattern for $\phi \neq 0$ arises due to the 
      appearance of multiple strong peaks in the electric field. \txt{Left:}{Top:} The spectrum for configurations with $\phi=\pi/4$ and $\phi=3\pi/4$ appear
      on top of each other. \txt{Right:}{Bottom:} Symmetry in $p_z$ is broken, because we take into account only one spin direction within a DHW calculation.
      Parameters: $\varepsilon=0.5$, $\tau=10$ $m^{-1}$, $\omega=0.1$ $m$ and $\lambda=10 ~ m^{-1}$.}
      \label{fig:T10L10PxPz}
      \end{figure} 

Interference patterns and a shift towards higher momentum $p_z$ also show up in the distributions $n \left( p_x \right)$ and $n \left( p_z \right)$, see Eq.~\eqref{equ:nn}.
Fig.~\ref{fig:T10L10PxPz} serves as an example, where we have integrated out one momentum coordinate, respectively. We used the same parameters as in Fig.~\ref{fig:T10L10};
$\varepsilon=0.5$, $\tau=10$ $m^{-1}$, $\omega=0.1$ $m$ and $\lambda=10$ $m^{-1}$.

Analyzing the particle distribution as a function of the momentum $p_x$ we observe remarkable agreement with results obtained within the QKT framework \cite{Hebenstreit:2009km}.
This is insofar astonishing as firstly QKT describes pair creation in purely time-dependent electric fields, while we have taken spatial inhomogeneities as well as
a strong magnetic field into account. Secondly, the particle density in the whole momentum space does not look familiar at all, see Fig.~\ref{fig:T10L10}.

Nevertheless, the function $n \left( p_x \right)$ displays all essential features of Schwinger pair production. Particles created within a single-peak electric field
show a smooth distribution function. In case of multiple peaks, however, the electric field acts as if it were a double-slit experiment in time. Another interesting
aspect, that still holds for moderately varying fields $\lambda=10$ $m^{-1}$, is given by the distributions $\phi=\pi/4$ and $\phi=3\pi/4$. Although their representation
in full momentum space  $\left( p_x, p_z \right)$ is different, their distributions as functions of $p_x$ cover each other.
This could be related to the fact, that both field configurations still possess the same field energy as well as the same general structure. To be more specific,
the relation $\mathbf{A} \left(t,z \right) \bigg|_{\phi=\pi/4} = \mathbf{A} \left(-t,z \right) \bigg|_{\phi=3\pi/4}$ holds.       

A completely different picture is drawn by the function $n \left( p_z \right)$. Not only is the link between field configurations of type $\phi=\pi/4$ and $\phi=3\pi/4$
not visible, also any obvious signature of an interference pattern is integrated out. 
Nevertheless, all configurations display a shift to higher momentum $p_z$, although the strength of the effect varies. 
     
\section{Summary}

Based on numerical solutions within the DHW approach we have discussed the Schwinger pair production process in spatiotemporally inhomogeneous few-cycle background fields.
The DHW formalism provides access to all phase-space informations. Employing advanced numerical methods, we have been able to compute particle momentum spectra as well as 
spatial-momentum distribution functions in order to thoroughly investigate how spatial and temporal variations in the electric and magnetic fields affect the particle distribution. 
Furthermore, we have introduced a semi-classical model on the basis of an effective theory for the particle production rate supplemented by a rigorous trajectory analysis.
This model served as a supporting tool providing an additional point of view and facilitating our interpretation of time-resolved Schwinger pair production.

Our main goal was to investigate particle creation in the vicinity of an additional strong and inhomogeneous magnetic field. 
We have found remarkable signatures of quantum interferences and spin-field interactions. Additionally, we observed the formation of characteristic patterns
strongly depending on the carrier-envelope phase of the background fields.
To sum up, the inhomogeneous magnetic field turns out to be a decisive factor towards understanding pair production under realistic conditions. 

We have introduced various strategies enabling us to perform calculations within a phase-space formalism without any additional truncations opening up the possibility
to perform calculations for more realistic field configurations. Correspondingly, the trajectory-based model can be easily extended to more advanced field configurations, too. 
Moreover, due to the fact that one has full control of the particles in the semi-classical model, it should be possible to expand it such that one can take electron-electron interactions
as well as radiation reaction effects into account.
The inclusion of phase information is conceptually more difficult considering that one probably wants to keep the simple and easy-to-use form of the approach. 
Nevertheless, a combination of both methods appears to be promising particularly with regard to future challenges in the research field.

\begin{acknowledgements}

We want to thank Holger Gies, Alexander Blinne and Andr\'{e} Sternbeck for many fruitful discussions.
We are very grateful to Holger Gies for comments on the manuscript.
The work of CK is funded by the BMBF under grant No.\  05P15SJFAA (FAIR-APPA-SPARC) and by the
Helmholtz Association through
the Helmholtz Postdoc Programme (PD-316).

Computations were performed on the ``Supermicro Server 1028TR-TF'' in Jena, which
was funded by the Helmholtz Postdoc Programme (PD-316).

\end{acknowledgements}

\end{document}